\documentclass[fleqn,usenatbib]{mnras}

\usepackage{newtxtext,newtxmath}
\usepackage[T1]{fontenc}

\DeclareRobustCommand{\VAN}[3]{#2}
\let\VANthebibliography\thebibliography
\def\thebibliography{\DeclareRobustCommand{\VAN}[3]{##3}\VANthebibliography}

\usepackage{graphicx}	% Including figure files
\usepackage{amsmath}	% Advanced maths commands
\usepackage{siunitx}

\title[LEGA-C gradients]{Age and metal gradients in massive quiescent galaxies at $0.6 \lesssim z \lesssim 1.0$: implications for quenching and assembly histories} 

\author[C. M. Cheng et al.]{
Chloe M. Cheng,$^{1}$\thanks{E-mail: cheng@strw.leidenuniv.nl (CMC)}
Mariska Kriek,$^{1}$
Aliza G. Beverage,$^{2}$
Arjen van der Wel,$^{3, 4}$
Rachel Bezanson,$^{5}$
\newauthor
Francesco D'Eugenio,$^{6,7}$
Marijn Franx,$^{1}$
Pavel E. Mancera Piña,$^{1}$
Angelos Nersesian,$^{3, 8}$
Martje Slob,$^{1}$
\newauthor
Katherine A. Suess,$^{9}$
Pieter G. van Dokkum,$^{10}$
Po-Feng Wu,$^{11,12,13}$
Anna Gallazzi,$^{14}$
and Stefano Zibetti$^{14}$
\\
$^{1}$Leiden Observatory, Leiden University, P.O. Box 9513, 2300 RA Leiden, The Netherlands\\
$^{2}$Department of Astronomy, University of California, Berkeley, CA 94720, USA\\
$^{3}$Sterrenkundig Observatorium, Universiteit Ghent, Krijgslaan 281 S9, B-9000 Gent, Belgium\\
$^{4}$Max-Planck-Institut für Astronomie, Königstuhl 17, D-69117, Heidelberg, Germany\\
$^{5}$Department of Physics \& Astronomy and PITT PACC, University of Pittsburgh, Pittsburgh, PA 15260, USA\\
$^{6}$Kavli Institute for Cosmology, University of Cambridge, Madingley Road, Cambridge, CB3 0HA, United Kingdom\\
$^{7}$Cavendish Laboratory - Astrophysics Group, University of Cambridge, 19 JJ Thomson Avenue, Cambridge, CB3 0HE, United Kingdom\\
$^{8}$STAR Institute, Université de Liège, Quartier Agora, Allée du Six Août 19c, 4000 Liège, Belgium\\
$^{9}$Kavli Institute for Particle Astrophysics and Cosmology and Department of Physics, Stanford University, Stanford, CA 94305, USA\\
$^{10}$Astronomy Department, Yale University, 52 Hillhouse Ave, New Haven, CT 06511, USA\\
$^{11}$Institute of Astrophysics, National Taiwan University, Taipei 10617, Taiwan\\
$^{12}$Department of Physics and Center for Theoretical Physics, National Taiwan University, Taipei 10617, Taiwan\\
$^{13}$Physics Division, National Center for Theoretical Sciences, Taipei 10617, Taiwan\\
$^{14}$INAF-Osservatorio Astrofisico di Arcetri, Largo Enrico Fermi 5, I-50125 Firenze, Italy
}

\date{Accepted 2024 July 12. Received 2024 July 10; in original form 2024 May 27}

\pubyear{2024}

\begin{document}
\label{firstpage}
\pagerange{\pageref{firstpage}--\pageref{lastpage}}
\maketitle

\begin{abstract}
We present spatially resolved, simple stellar population equivalent ages, stellar metallicities, and abundance ratios for 456 massive ($10.3\lesssim\log(\mathrm{M}_*/\mathrm{M}_\odot)\lesssim11.8$) quiescent galaxies at $0.6\lesssim z\lesssim1.0$ from the Large Early Galaxy Astrophysics Census, derived using full-spectrum models.  Typically, we find flat age and [Mg/Fe] gradients, and negative [Fe/H] gradients, implying iron-rich cores.  We also estimate intrinsic [Fe/H] gradients via forward modelling.  We examine the observed gradients in three age bins.  Younger quiescent galaxies typically have negative [Fe/H] gradients and positive age gradients, possibly indicating a recent central starburst.  Additionally, this finding suggests that photometrically measured flat colour gradients in young quiescent galaxies are the result of the positive age and negative metallicity gradients cancelling each other.  For older quiescent galaxies, the age gradients become flat and [Fe/H] gradients weaken, though remain negative.  Thus, negative colour gradients at older ages are likely driven by metallicity gradients.  The diminishing age gradient may result from the starburst fading.  Furthermore, the persistence of the [Fe/H] gradients may suggest that the outskirts are simultaneously built up by mergers with lower metallicity satellites.  On the other hand, the gradients could be inherited from the star-forming phase, in which case mergers may not be needed to explain our findings.  This work illustrates the need for resolved spectroscopy, instead of just photometry, to measure stellar population gradients.  Extending these measurements to higher redshift is imperative for understanding how stellar populations in quiescent galaxies are assembled over cosmic time.
\end{abstract}


%%%%%%%%%%%%%%%%%%%%%%%%%%%%%%%%%%%%%%%%%%%%%%%%%%

%%%%%%%%%%%%%%%%% BODY OF PAPER %%%%%%%%%%%%%%%%%%

\section{Introduction}\label{sec:introduction}
In the low-redshift ($z\sim0$) Universe, spatially resolved stellar populations have given us many insights into how nearby galaxies formed.  In particular, we have learned about the assembly histories of massive, quiescent galaxies by studying their stellar populations out to large radii.  These radial stellar population gradients encode the build-up of stellar mass and allow us to differentiate quenching mechanisms and assembly scenarios (see \citealt{Maiolino_2019} for a review). 

It has been found that low-$z$ quiescent galaxies have negative radial colour gradients, with redder centres and bluer outskirts (i.e. \citealt{Peletier_1989, Franx_1990, Peletier_1990b, Peletier_1990a, Saglia_2000, La_Barbera_2005, Sun_2010, Tortora_2010, Gonzalez_Perez_2011, Parikh_2021, Liao_2023}).  In addition, their $\alpha$-element abundances are generally consistent with having no gradients, they have flat or mildly positive gradients in age (i.e. galaxy centres are the same age or slightly younger than their outskirts, though note that, e.g., \citealt{Zibetti_2020} found U-shaped age profiles), and they have mildly negative gradients in [Fe/H] (i.e. galaxy centres are more metal-rich than their outskirts, \citealt{Mehlert_2003, Rose_2005, Kuntschner_2006, Sanchez_Blazquez_2007, Koleva_2011, Greene_2013, Greene_2015, Greene_2019, Pastorello_2014, Gonzalez_Delgado_2015, Cook_2016, Goddard_2017, Martin_Navarro_2018, San_Roman_2018, Ferreras_2019, Oyarzun_2019, Lacerna_2020, Zheng_2019, Santucci_2020, Lee_2023, Yoon_2023, Parikh_2024}, etc.).  Thus, colour gradients are thought to be primarily driven by metallicity gradients, with perhaps some contribution from age gradients \citep{Peletier_1989, Franx_1990, Peletier_1990a, Davies_1993, Vazdekis_1997, Saglia_2000, La_Barbera_2005, Sun_2010, Tortora_2010, Parikh_2021, Liao_2023, Miller_2023b}.  These findings are indicative of inside--out growth, either because younger and/or lower metallicity stellar populations at the outskirts are accreted by minor mergers (e.g. \citealt{Naab_2009}), or because late-time star formation occurs in the disc and not in the central bulge (i.e. \citealt{Abramson_2014, Hill_2017}).

While these findings have revealed a great deal about the assembly histories of nearby galaxies, low-$z$ studies do not paint the entire picture.  For example, if there has been a significant amount of merging in a galaxy's past, then the stars that we see today were born in a range of different progenitor galaxies with potentially very different formation histories.  Additionally, radial migration can occur with time, so present-day stars may not be in the same places as when they were born \citep{Maiolino_2019}.  Therefore, to study conditions in the main progenitors of today's elliptical galaxies and to constrain past merging activity, we need to observe quiescent galaxies closer to the epochs of formation and quenching, before the majority of this merger activity occurred.  

At higher $z$, measurements of resolved stellar populations of quiescent galaxies have been almost exclusively limited to photometry.  For example, \cite{Wuyts_2010}, \cite{Guo_2011}, \cite{Szomoru_2012}, \cite{Chan_2016}, \cite{Ciocca_2017}, \cite{Liu_2017}, \cite{Mosleh_2017}, \cite{Suess_2019, Suess_2019b, Suess_2020, Suess_2021}, and \cite{Miller_2023b} showed that quiescent galaxies between $0.5 \lesssim z \lesssim 2.5$ have negative colour gradients on average, similar to low-$z$ results.  However, it is not yet clear what physical property is driving the observed colour gradients at these redshifts.  In particular, gradients in age, metallicity, and dust are possible contenders, but it is difficult to disentangle these effects beyond $z \sim 0$ (\citealt{Suess_2019}).  For example, \cite{Gargiulo_2012} found that the negative colour gradients in several galaxies at $z\sim1.5$ could be explained either by a pure radial age \textit{or} metallicity variation.  Thus, complementary studies using high signal-to-noise (S/N) spectroscopy are needed to understand detailed star formation and assembly histories. 

Unfortunately, spectroscopic studies of quiescent galaxies beyond the low-$z$ universe are exceedingly challenging.  Therefore, they have largely been based on the integrated light of distant quiescent galaxies.  In general, these studies have found that quiescent galaxies tend to have extreme elemental abundances, with metal-poor and $\alpha$-enhanced stellar populations compared to local galaxies (i.e. \citealt{Kriek_2016, Kriek_2019, Jafariyazani_2020, Jafariyazani_2024, Beverage_2021, Beverage_2023, HM_Beverage, Carnall_2022, Zhuang_2023}).  However, these findings from integrated light are biased toward the central regions of quiescent galaxies.  In order to truly understand the evolution that took place, we require spatially resolved spectroscopic measurements of a statistically significant sample of distant quiescent galaxies.  

Thus far, this has not been possible due to observational limitations, as we require high-quality spectra of faint absorption lines, which are shifted to the near-infrared beyond $z = 1$.  As a result, spectroscopic measurements of spatially resolved stellar populations have only been achieved for a handful of individual distant systems.  For example, \cite{Jafariyazani_2020} examined a massive, lensed galaxy at $z \sim 2$ and found that, similar to galaxies at low-$z$, this galaxy has no age or [Mg/Fe] gradients and a marginally negative [Fe/H] gradient.  \cite{Akhshik_2023} examined eight lensed galaxies using \textit{Hubble Space Telescope (HST)} grism spectroscopy and found diverse age gradients.  They also measured metallicity gradients, but due to the low spectral resolution their uncertainties were extremely large.  Finally, \cite{Perez_Gonzalez_2024} used \textit{James Webb Space Telescope (JWST)}/Near Infrared Spectrograph (NIRSpec) integral field unit (IFU) data to measure a strongly negative total metallicity gradient in a massive quiescent galaxy at $z\sim3.7$.

To bridge this gap, we therefore need to resolve stellar populations in larger samples of distant quiescent galaxies.  To achieve this goal, we require ultradeep spectra with at least moderate spatial resolution, so that multiple resolution elements can be analysed.  We also require a large, statistically significant sample of quiescent galaxies.  The Large Early Galaxy Astrophysics Census (LEGA-C, \citealt{van_der_Wel_2016, van_der_Wel_2021, Straatman_2018}), a high S/N, high spectral resolution spectroscopic survey of 3600 galaxies at $0.6\lesssim z \lesssim 1$, has finally made this possible.  In this work, we present a comprehensive study of spatially resolved stellar population parameters in a statistically significant sample of quiescent galaxies at $0.6\lesssim z \lesssim 1$, using data from the third data release of LEGA-C.

This paper is organized as follows: in Section~\ref{sec:data_sample}, we describe the LEGA-C data and our sample selection.  In Section~\ref{sec:methods}, we outline our spectral extraction, considerations for the blurred nature of the data due to observational seeing, and full-spectrum stellar population synthesis (SPS) modelling.  We present our results and our predictions for gradients in intrinsic space in Section~\ref{sec:results}.  We discuss the implications of our results for photometric measurements in the literature, galaxy assembly scenarios, and our assumptions in Section~\ref{sec:discussion}.  Finally, our conclusions are presented in Section~\ref{sec:conclusions}.

\begin{figure*}
    \centering
    \includegraphics[width=\textwidth]{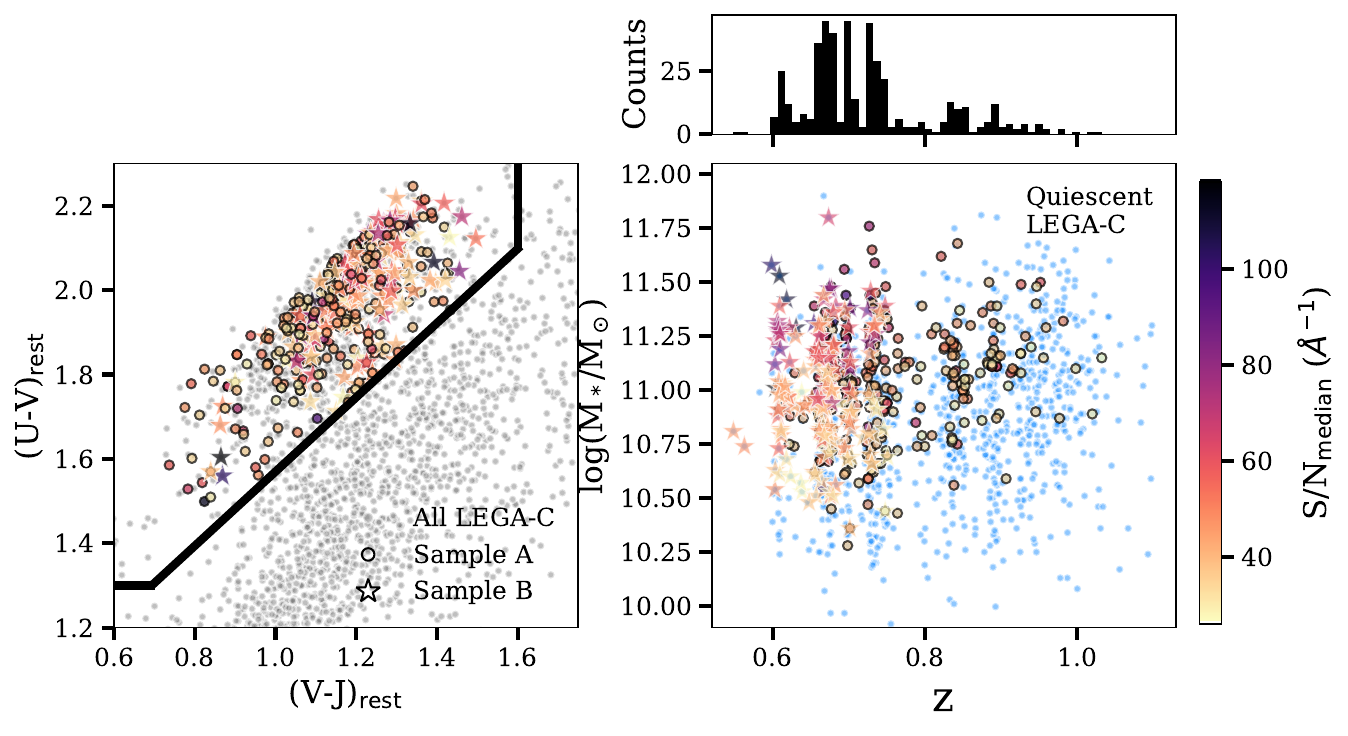}
    \caption{Quiescent galaxy sample studied in this work.  In the left panel, we show the rest-frame $UVJ$ diagram.  The full LEGA-C sample is shown as grey points.  In the right panel, we show stellar mass as a function of redshift.  The full quiescent sample in LEGA-C is shown as blue points.  In each panel, symbols are colour coded by the median S/N of the integrated spectrum.  Circles represent galaxies for which we can only measure spatially resolved age and [Fe/H] (Sample A).  Stars represent galaxies for which we can measure spatially resolved age, [Fe/H], and [Mg/Fe] (Sample B).  A histogram of the redshifts of the galaxies in Sample B is shown in the top panel.}
    \label{fig:paper_sample}
\end{figure*}

\begin{figure}
    \centering
    \includegraphics[width=\columnwidth]{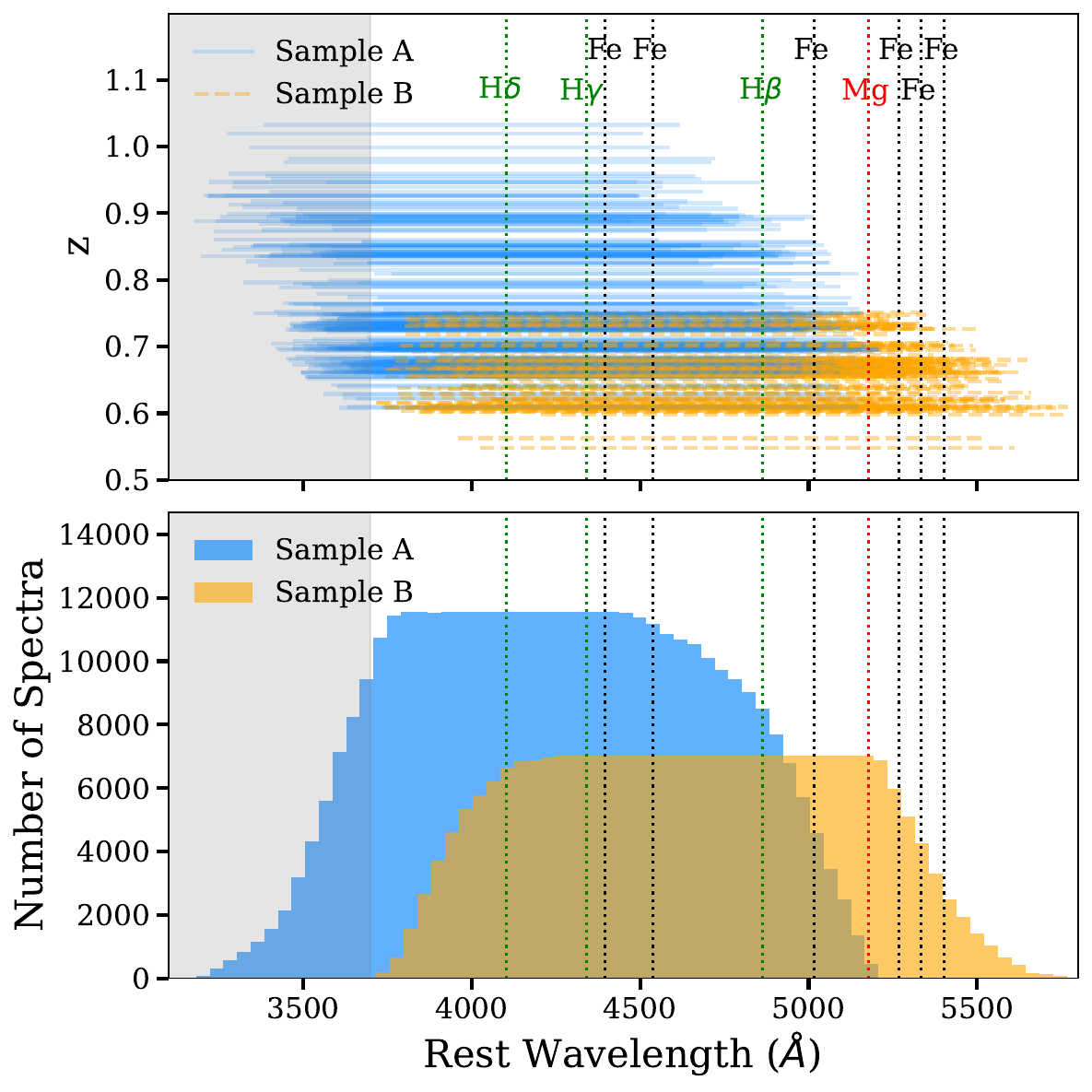}
    \caption{The wavelength coverage of our sample.  In each panel, Sample A is indicated in blue and Sample B is indicated in orange.  The shaded region indicates wavelengths that cannot be fit with \textsc{alf}.  We indicate key Balmer, Fe, and Mg features with vertical dotted lines. \textit{Top}: the total wavelength ranges covered by all of the spectra in our sample.  Rest wavelength is shown on the $x$-axis and redshift is shown on the $y$-axis.  Galaxies in Sample A (for which we can measure age and metallicity gradients) are shown as solid lines and galaxies in Sample B (for which we can also measure Mg gradients) are shown as dashed lines.  \textit{Bottom}: histograms of the wavelength coverage of Samples A and B.}
    \label{fig:waveranges}
\end{figure}

\section{Data and sample}\label{sec:data_sample}
\subsection{LEGA-C spectra}\label{sec:lega-c}
In this work we use spectroscopic data from the third data release of LEGA-C, a European Southern Observatory (ESO) Public Spectroscopic survey of 3600 galaxies between $0.6 \lesssim z \lesssim 1.0$.  These galaxies are located in the Cosmic Evolution Survey (COSMOS) field \citep{Scoville_2007} and were selected from the UltraVISTA \citep{McCracken_2012} \textit{K}-band catalogue by \cite{Muzzin_2013}.  The data were collected over 128 nights using the VIsible MultiObject Spectrograph (VIMOS) on the ESO Very Large Telescope (\textit{VLT}), providing deep (20-h integration), $R\sim3500$ spectra with an average S/N $\sim20$ \AA$^{-1}$.  The two-dimensional (2D) and reduced one-dimensional (1D) spectra are available on the ESO Science Archive Facility\footnote{\url{http://archive.eso.org/eso/eso_archive_main.html}~.}\textsuperscript{,}\footnote{The reduced 1D spectra and catalogue have been released by ESO (\url{http://archive.eso.org/cms/eso-archive-news/Third-and-final-release-of-the-Large-Early-Galaxy-Census\\-LEGA-C-Spectroscopic-Public-Survey-published.html}), and are also available here: \url{https://users.ugent.be/~avdrwel/research.html\#legac}~.}.  See \cite{van_der_Wel_2016, van_der_Wel_2021}, and \cite{Straatman_2018} for details.  In this study, we use the reduced 2D spectra.

\subsection{Sample selection}\label{sec:sample_selection}
We select quiescent galaxies from the LEGA-C catalogue via their rest-frame $U-V$ and $V-J$ colours \citep{Wuyts_2007}.  To make this selection, we first determine the UltraVISTA colours \citep{Muzzin_2013} for the LEGA-C galaxies with \textsc{eazy} \citep{EAZY} by fixing the spectroscopic $z$ to those measured from the LEGA-C spectra \citep{van_der_Wel_2021}.  We use these colours to characterize galaxies as star forming or quiescent, employing the classification from \cite{Muzzin_2013}, and select the quiescent galaxies.  We show the rest-frame $UVJ$ diagram in the left panel of Fig.~\ref{fig:paper_sample}, where galaxies in our sample are colour coded by their median rest-frame S/N per \AA.  We remove galaxies with spectroscopic $z < 0.5$ from our sample.  We also discard galaxies in LEGA-C's mask 2, as we find that the noise spectrum is significantly underestimated.  Furthermore, we require that each spectrum that we examine has a median rest-frame S/N $\gtrsim20$ \AA$^{-1}$ and a maximum wavelength of at least $4450$ \AA\ for us to sufficiently recover ages and metallicities (these limits are determined using simulated galaxies, see Section~\ref{sec:alf} and Appendix~\ref{sec:appendix_mocks}).  We perform a final sample cleaning during the fitting stage and discard individual galaxies whose age, [Z/H], [Fe/H], and [Mg/H] posterior distributions are not well sampled.  We also examine all ages and ensure that they are $\geq 1$ Gyr as our fitting code is only valid for stellar populations older than 1 Gyr.  See Section~\ref{sec:alf} for details about our fitting procedure.  

We split our sample into two subsamples.  For each galaxy, we make use of the full available wavelength range.  For those galaxies with a maximum wavelength $\geq5207$ \AA\ (i.e. galaxies which include at least the bluest Mg feature, Mgb), we are able to measure age, metallicity, and Mg gradients (Sample B).  However, for those galaxies whose spectra have wavelengths anywhere between $3700$ \AA\ $ \leq\lambda < 5207$ \AA\ (i.e. with no Mg features), we still expect to recover age and metallicity gradients (Sample A).  In particular, spectra in Sample A have a wavelength range of at least $3700.0$ \AA $\lesssim \lambda \lesssim 4468.9$ \AA\ which allows us to capture multiple Balmer and Fe lines.  We show the wavelength ranges encompassed by each galaxy in our sample in the top panel of Fig.~\ref{fig:waveranges}.  We indicate galaxies in Sample A with solid lines and galaxies in Sample B with dashed lines.  In the bottom panel, we show histograms of the total wavelength coverage of Sample A (blue) and Sample B (orange).  In Fig.~\ref{fig:paper_sample}, we show the galaxies in Sample A as circles and the galaxies in Sample B as stars.  In total, Sample A contains 285 galaxies while Sample B contains 171 galaxies (in other words, we can measure age and metallicity gradients for a total of 456 galaxies and Mg gradients for 171 galaxies).

Our final sample of 456 quiescent galaxies is shown in the right panel of Fig.~\ref{fig:paper_sample}, where spectroscopic redshift (from LEGA-C) is shown on the $x$-axis and stellar mass (computed by the LEGA-C team using \textsc{magphys}; \citealt{MAGPHYS}) is shown on the $y$-axis.  The distribution of redshifts is also shown in the top panel.  The rest of the quiescent LEGA-C sample is shown in blue.  Our selection covers a wide range of redshifts ($0.5 \leq z \leq 1.1$) and stellar masses ($10.28 \leq \log(\mathrm{M}_*/\mathrm{M}_\odot) \leq 11.80$).  We note that we include $\sim31$ per cent of the total quiescent sample in LEGA-C.  In the left panel of Fig.~\ref{fig:paper_sample}, we show that we sample the full distribution of quiescent galaxies in $UVJ$-space and our sample is thus representative of the $UVJ$ quiescent sequence.  

\section{Methods}\label{sec:methods}
To determine spatially resolved stellar population parameters, we extract spatially resolved spectra, determine observed sizes in convolved space, and fit the spectra with full-spectrum SPS models.  We describe our spectral extraction in Section~\ref{sec:extraction}, our size determination in Section~\ref{sec:convolved_re}, and our spectral fitting in Section~\ref{sec:alf}.

\subsection{Optimal extraction}\label{sec:extraction}
\begin{figure*}
    \centering
    \includegraphics[width=\textwidth]{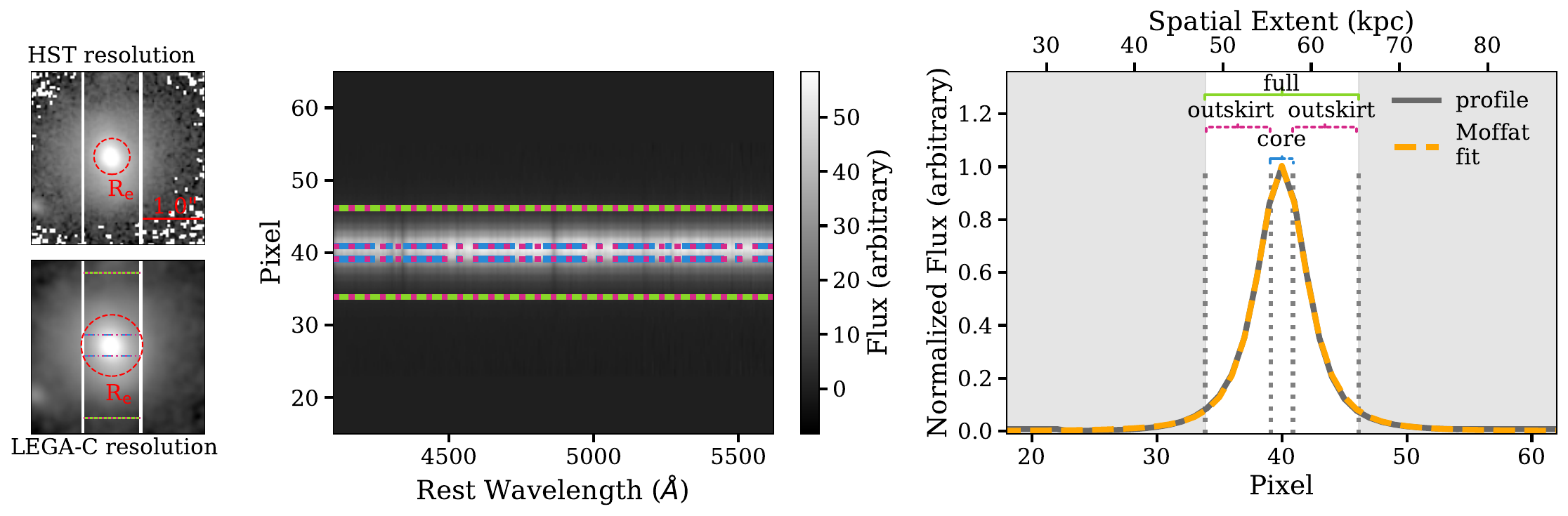}
    \caption{\textit{Left}: COSMOS \textit{HST} ACS images \citep{COSMOS_ACS_mosaics, Scoville_2007} of an example galaxy.  The top panel is the image at \textit{HST}/\textit{F814W} resolution, with the VIMOS slit overplotted.  The bottom panel is the same image but convolved to the seeing of LEGA-C, which we compute in Section~\ref{sec:convolved_re}.  We also outline the pixels that are included in each extraction, where solid lines show the full, integrated extraction region, dotted lines show the outskirt region, and dashed--dotted lines show the core region.  The $R_{\rm e}$ is shown at the \textit{HST} and LEGA-C resolutions in each panel and is indicated by the dashed circles.  The remaining two panels show a demonstration of our optimal extraction procedure.  In each panel, we outline the extraction regions similarly.  \textit{Middle}: an example 2D spectrum for the same galaxy as in the HST images.  \textit{Right}: the collapsed normalized flux profile of this 2D spectrum (solid curve), fit with a Moffat profile (dashed curve).  The extraction regions are indicated with labels and dotted lines.  Grey shaded regions show the pixels not included in our extracted spectra.  On the top $x$-axis, we use the pixel scale of VIMOS (\SI{0.205} arcsec) to transform the pixels to physical units.}
    \label{fig:paper_extraction}
\end{figure*}

\begin{figure*}
    \centering
    \includegraphics[width=\textwidth]{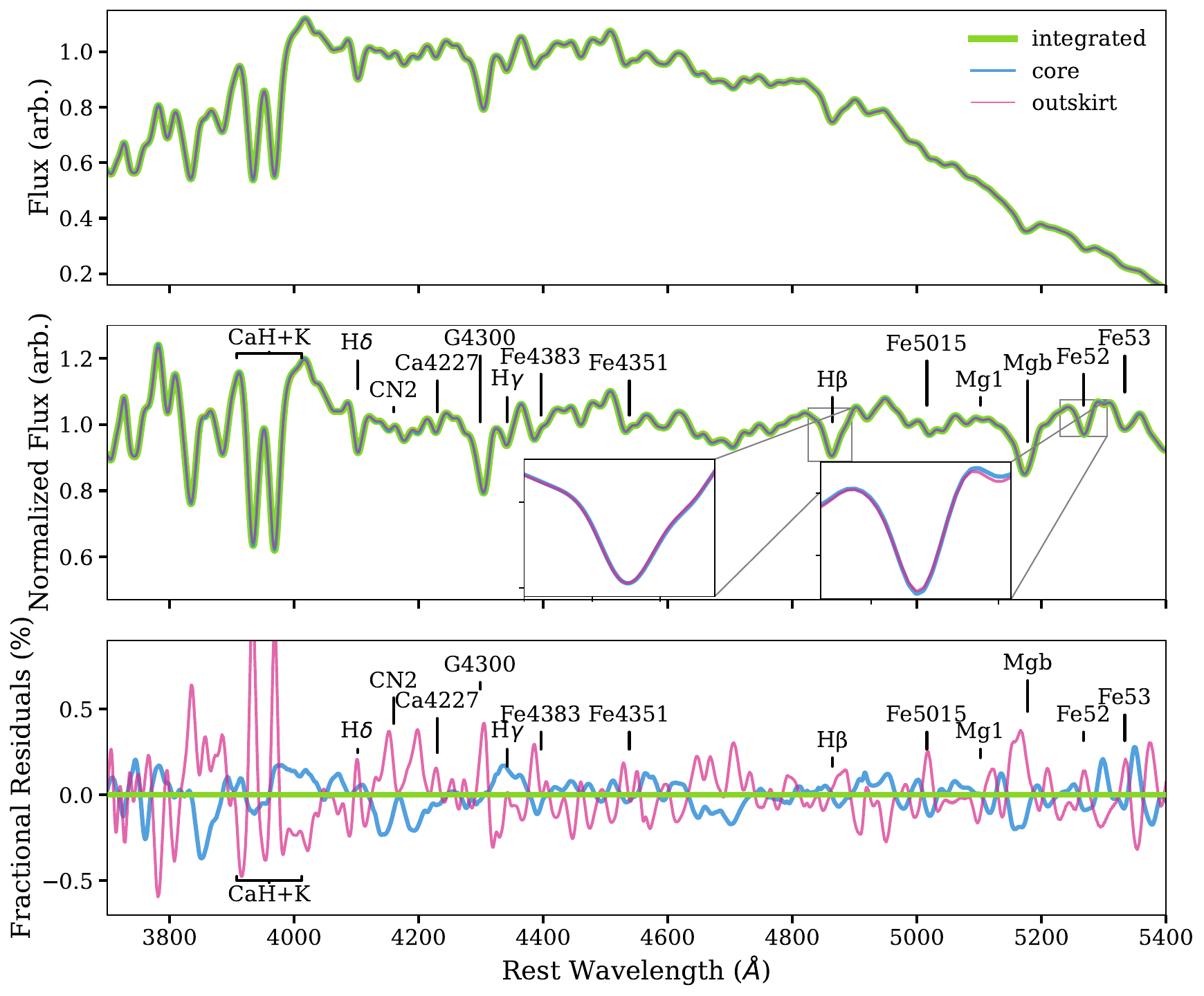}
    \caption{\textit{Top}: extracted and stacked integrated, core, and outskirt spectra from our entire sample.  We smooth each spectrum to a common velocity dispersion of $\sigma = 450$ km s$^{-1}$, then mean stack the spectra from each extraction.  \textit{Middle}: the same spectra but continuum normalized.  We label key spectral absorption features.  In the inset panel, we zoom-in on the core and outskirt spectra near the H$\beta$ feature, which is sensitive to age, and near the Fe52 feature, which is sensitive to [Fe/H].  The two spectra are distinct in Fe52, but not in H$\beta$.  \textit{Bottom}: a representation of the fractional residuals of each stacked spectrum, where we divide each spectrum by the integrated spectrum.}
    \label{fig:paper_spectra}
\end{figure*}

To measure spatially resolved stellar population parameters in the LEGA-C galaxies, for each galaxy we obtain 1D spectra for two spatial bins and a 1D integrated spectrum using a custom optimal extraction routine applied to the 2D spectra.

For our optimal extraction routine, we first obtain the flux profile of each galaxy by collapsing the 2D spectrum over the wavelength axis.  We then fit a \cite{Moffat_1969} profile to the flux profile, which is given by
\begin{equation}\label{eq:moffat}
    M(\alpha, \gamma) = A\left(1 + \frac{(x - x_0)^2}{\gamma^2}\right)^{-\alpha}.
\end{equation}
Here, we fit for $A$ (a normalization factor), $x_0$ (the centre of the profile), and $\gamma$ and $\alpha$ (the Moffat parameters).  The full width at half-maximum (FWHM) of the Moffat profile is given by $FWHM = 2\gamma\sqrt{2^{1/\alpha} - 1}$.  Moffat profiles were also used to extract the public LEGA-C spectra \citep{van_der_Wel_2021}, and have been found to fit the wings of spectral profiles better than Gaussian fits \citep{Moffat_1969, Trujillo_2001}.

We extract three 1D spectra for each galaxy: an integrated spectrum and two spatially binned spectra.  To extract an integrated spectrum, we weight all (sub-)rows in the 2D spectrum by the Moffat profile and sum the weighted (sub-)rows with significant flux [i.e. (sub-)rows between the 3rd and 97th percentiles of the Moffat profile].  For the spatially binned spectra, we perform a similar weighted sum of the (sub-)rows in the middle 30 per cent of the Moffat profile only.  These comprise the `core' of the galaxy.  To extract the `outskirts', we perform a weighted sum of the remaining (sub-)rows (excluding the 3rd and 97th percentiles).  A demonstration of this procedure is shown in Fig.~\ref{fig:paper_extraction}.  We define these extraction regions to compromise between a large sample size and a minimum difference in S/N between each core and outskirt spectrum.  In particular, the S/N of the core and outskirt spectra for each galaxy are often comparable.  However, we have tested different sizes of extraction regions and find that the exact combination of rows that are included does not affect our conclusions.  See Section~\ref{sec:caveats} for more details. 

For visualization purposes, we stack all of our rest-frame integrated, core, and outskirt spectra, respectively.  Prior to stacking, we regrid the spectra to the same wavelength array using \textsc{spectres} \citep{SPECTRES}, continuum-normalize them, and smooth them all to a common velocity dispersion of $\sigma = 450$ km s$^{-1}$ (the maximum $\sigma$ of our sample).  In Fig.~\ref{fig:paper_spectra}, we show these spectra as well as their fractional residuals (where we divide each stacked spectrum by the stacked integrated spectrum).  Visually, it can be seen that the three spectra are distinct.  In particular, in the middle panel, we zoom in on two strong features: H$\beta$, which is sensitive to age, and an Fe feature near $5200$ \AA, which is sensitive to [Fe/H].  The H$\beta$ feature is perhaps slightly weaker for the core spectrum.  However, the Fe feature is distinctly stronger for the core spectrum compared to the outskirt spectrum.  We discuss this further in Section~\ref{sec:results}.

\subsection{Convolved $R_{\mathrm{e}}$}\label{sec:convolved_re}
A major consideration of our spectral extraction method is that our measurements will be significantly affected by the seeing of the telescope.  In other words, our data and therefore our measurements are blurred by instrumental and atmospheric effects.  It would therefore be a misrepresentation to display our results in units of the $R_{\mathrm{e}}$ measured in the \textit{HST}/\textit{F814W} images \citep{Scoville_2007} via \textsc{galfit} \citep{galfit} by the LEGA-C team.  As such, we derive the \textit{convolved} $R_{\rm e}$ for all galaxies using a method similar to \cite{Price_2016}, where the $R_{\mathrm{e}}$ is similarly blurred by the seeing of the observations.  

We first determine the seeing by creating an idealized mock galaxy image in \textsc{galfit} \citep{galfit} for each galaxy in our sample, using the structural parameters reported in LEGA-C.  We generate the images with the same spatial resolution as LEGA-C, where we create square pixels of $0.205$ arcsec by $0.205$ arcsec, and with the same number of pixels as the LEGA-C spectra.  We convolve this model image with a grid of different model point spread functions (PSFs).  Each model PSF is a 2D Moffat kernel with different $\alpha$ and $\gamma$ parameters, with $\alpha$ and $\gamma$ ranging between 0.1 - 7.0 (see equation~\ref{eq:moffat}).  We then mask each model image with a rectangular aperture the same size as the VIMOS slit ($1$ arcsec wide) using the \textsc{photutils} package \citep{photutils}.  We sum over the slit width to obtain a model flux profile ($m$).  We centre $m$ to the flux profile from the spectrum ($f$).  We normalize $m$ to $f$ by multiplying $m$ by a scaling factor defined as $\frac{\sum\left(m\times f\right)}{\sum m^2}$ (where we sum over all spatial elements) and compare it to $f$ by calculating the reduced $\chi^2$ ($\chi^2_{\rm red}$) value over the region that we consider in the optimal extraction (i.e. excluding the low-flux edges of the profiles).  The seeing value which produces the minimum $\chi^2_{\rm red}$ corresponds to our best-fit seeing.  Our median best-fit seeing is $\sim0.70$ arcsec and, as expected, we find a similar seeing within each LEGA-C mask.  We note that a similar method to determine the seeing in LEGA-C was used in \cite{van_Houdt_2021}, although there \textit{HST} Advanced Camera for Surveys (ACS)/\textit{F814W} images were convolved with the best-fit seeing instead of \textsc{galfit} models.  It is encouraging that our typical seeing is similar to that measured in \cite{van_Houdt_2021} ($0.75$ arcsec).

To calculate the convolved $R_{\rm e}$ for the LEGA-C spectra, we apply this best-fit seeing value to a similar model image of each galaxy, but with an increased spatial resolution ($0.1$ arcsec pixel$^{-1}$) and a larger number of pixels (50 times larger than the original model image).  We place a series of circular apertures of increasing size on the enlarged image and perform aperture sums within each circle using \textsc{photutils}.  We then determine the radius which encloses 50 per cent of the light for this convolved model.  For each galaxy, we use this value to determine the spatial extent of our core and outskirt spectra in units of this convolved $R_{\mathrm{e}}$.

\begin{figure*}
    \centering
    \includegraphics[width=\textwidth]{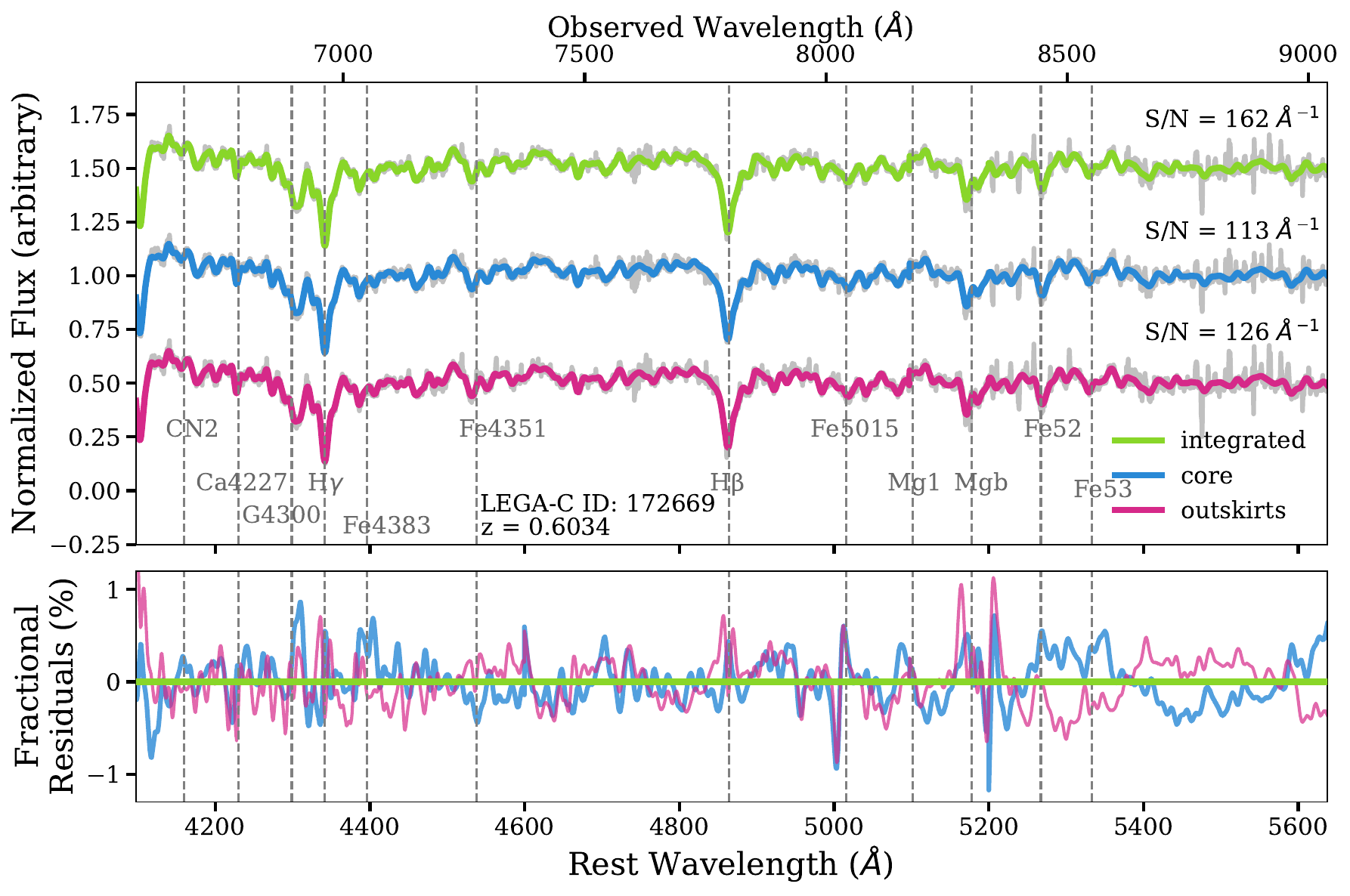}
    \caption{\textit{Top:} example best-fitting \textsc{alf} models to a $z\sim 0.6$ galaxy from our sample with high S/N. The spectra are shown in grey.  The integrated fit is shown on top, the fit to the core spectrum is shown in the middle, and the fit to the outskirt spectrum is shown at the bottom.  The median S/N of each bin is quoted above the respective fit.  Key spectral features are labelled in grey.  \textit{Bottom}: similar to Fig.~\ref{fig:paper_spectra}.  The fractional residuals of each fit, where we divide each fit by the integrated fit.}
    \label{fig:paper_fit}
\end{figure*}

\subsection{Full spectrum fitting}\label{sec:alf}
We fit each spectrum with the \textsc{absorption line fitter} (\textsc{alf}\footnote{\url{https://github.com/cconroy20/alf}}), a full spectrum SPS model \citep{CvD_2012a, Conroy_2018}, to derive our stellar population parameters.  The \textsc{alf} models are built on empirical simple stellar populations, created using the Mesa Isochrones and Stellar Tracks (MIST, \citealt{MIST}) and the Spectral Polynomial Interpolator  (SPI, \citealt{Villaume_2017}).\footnote{\url{https://github.com/AlexaVillaume/SPI_Utils}~.}  We use the Medium Resolution INT Library Of Empirical Spectra (MILES, \citealt{Sanchez_Blazquez_2006}), the Extended Infrared Telescope Facility stellar library (E-IRTF, \citealt{Villaume_2017}), and a large sample of M-dwarf spectra \citep{Mann_2015} with SPI.  In this way, \textsc{alf} develops stellar spectra as a function of $T_{\mathrm{eff}}$, surface gravity, and metallicity from a data-driven model.

The empirical parameter space spans $-2.0\lesssim$ [Fe/H] $\lesssim 0.5$ and $3.9\lesssim\log{(T_{\mathrm{eff}}/{\rm K})}\lesssim 3.5$ and is set by the combined E-IRTF and \cite{Mann_2015} samples.  We additionally use a theoretical stellar library (C3K, see \citealt{CvD_2012a}) to ensure the quality of interpolation at the boundaries of the empirical parameter space.  The \textsc{alf} models allow for variable abundance patterns by differentially including theoretical element response functions.  To derive the [Mg/Fe] versus [Fe/H] relation, we use the Mg abundances for the MILES stellar library stars from \cite{Milone_2011}.

\textsc{alf} first continuum normalizes the target spectrum by multiplying it by a high-order polynomial.  It then samples the posteriors of 46 stellar parameters using a \textsc{fortran} implementation of the Markov Chain Monte Carlo algorithm (MCMC) \texttt{emcee} \citep{emcee}, allowing for arbitrary variation in stellar age and detailed elemental abundance patterns.  It fits for systematic parameters to characterize observed errors.  Note that \textsc{alf} can fit spectra between $3700 - 24000$ \AA\ and can be used for stellar populations that are older than 1 Gyr.  For details, see \cite{CvD_2012a} and \cite{Conroy_2018}.  

We fit each spectrum using 1024 walkers, $20 000$ burn-in steps, and a $1 000$ step production run.  We examine 500 MCMC chains per fit in our analysis.  In our implementation of \textsc{alf}, we assume a \cite{Kroupa} initial mass function and fit a single age.  We initialize the age of each galaxy by drawing a random value from a uniform distribution centred at 3 Gyr to avoid the \textsc{alf} walkers getting trapped at an unrealistically high age.  We also set the upper limit of the age prior to be the age of the Universe at each galaxy's redshift, plus 2 Gyr to allow for uncertainties.  We do not fit the hot star component.   

We first fit the integrated spectra, allowing for variation in all stellar parameters in \textsc{alf}.  For the spatially binned spectra (core and outskirts), we do not expect to be able to constrain all of the elemental abundances that \textsc{alf} can fit at this S/N (see Appendix~\ref{sec:appendix_mocks}).  Thus, to accurately constrain age, Fe, and Mg, we use the values of all other abundances from the integrated fits and fix these values in the fits to the spatially binned spectra.  We inspect each of the three fits for each individual galaxy to ensure that the posterior distributions for the parameters of interest (age, Fe, and sometimes Mg) are well-sampled.  For those galaxies where the posteriors run up against the priors, we discard these fits as their ages, metallicities, or Mg abundances are not reliable.  This is the case for $< 10$ per cent of our sample.  This produces our final sample size of 456 galaxies. 

There are several galaxies in LEGA-C that have been observed twice (111 galaxies in our sample).  In our analysis, we treat these observations as different objects and fit them individually.  However, we also use these objects to ensure the robustness of our fitting method, by comparing our results for the duplicate galaxies as well as to the results when we stack and fit the duplicate galaxies.  In general, the results are consistent.  For more details, see Appendix~\ref{sec:duplicates}.  

In the top panel of Fig.~\ref{fig:paper_fit}, we show example fits for one of our high S/N galaxies at $z\sim0.6$.  The best fit to the integrated spectrum is shown in green, the best fit to the core spectrum is shown in blue, and the best fit to the outskirt spectrum is shown in magenta.  In the bottom panel, we divide each fit by the integrated fit, similar to the bottom panel of Fig.~\ref{fig:paper_spectra}.  The core and outskirt fits are distinct from each other.    

We compare the results of our integrated fits to those of the overlapping galaxies fit in \cite{Beverage_2023} (not shown) and find that our results are entirely consistent, despite the differences in the sample selection (see \citealt{Beverage_2023}) and spectral extraction (see \citealt{van_der_Wel_2021}).  We will discuss the results of our integrated fits in more detail in a future study.

\section{Results}\label{sec:results}
In this section, we present the results of our \textsc{alf} fits (Section~\ref{sec:observed_gradients}).  We also develop a model to explore what our observed gradients are expected to look like in intrinsic space (Section~\ref{sec:intrinsic_gradients}).

\subsection{Observed gradients}\label{sec:observed_gradients}
\begin{figure*}
    \centering
    \includegraphics[width=\textwidth]{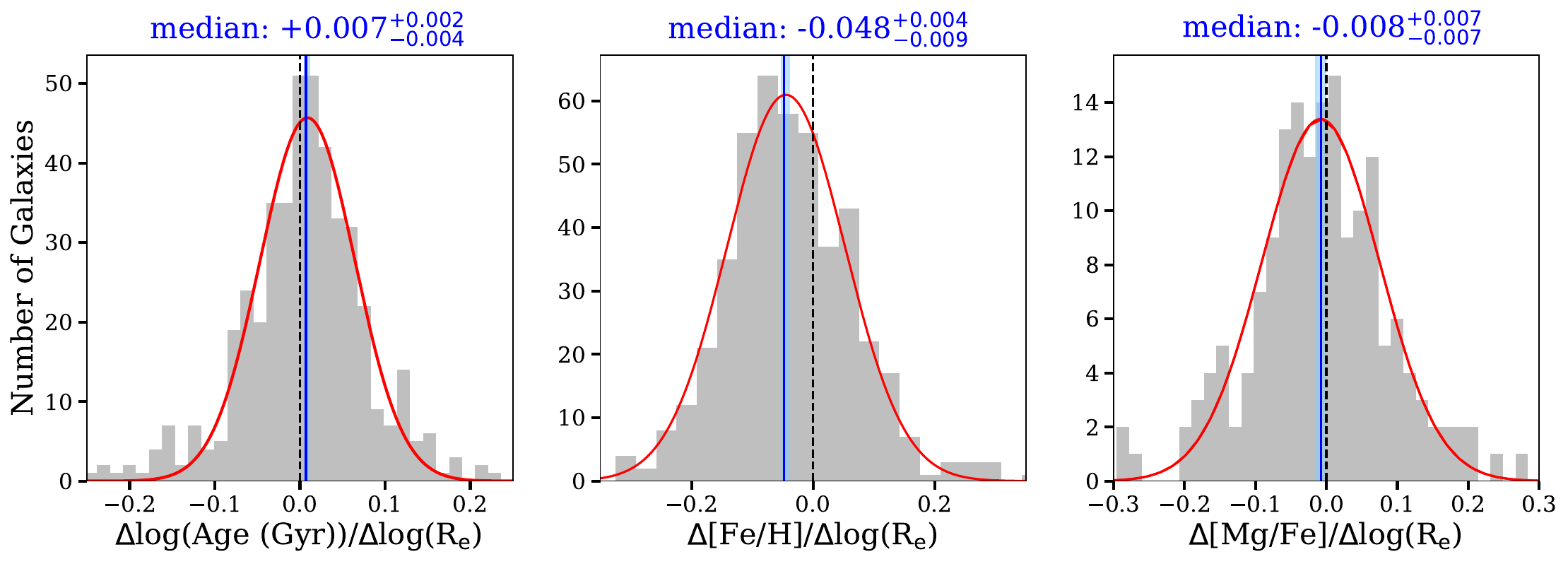}
    \caption{Histograms of our resolved stellar population gradients, normalized by $R_{\rm e}$ in convolved space.  Age gradients are shown on the left, metallicity gradients are shown in the middle, and [Mg/Fe] gradients are shown on the right.  In each panel, we fit the histogram of the gradients from our entire sample (grey) with Gaussians (red curves) to emphasize the shape of the distribution.  The median gradient is shown as a solid blue line, with the blue shaded region indicating the uncertainties on the median derived by bootstrap resampling.  We also show where a flat median gradient would lie (black dashed line).  Above each panel, we quote the value of the median gradient along with bootstrapped uncertainties.}
    \label{fig:grad_hist}
\end{figure*}

\begin{figure}
    \centering
    \includegraphics[width=\columnwidth]{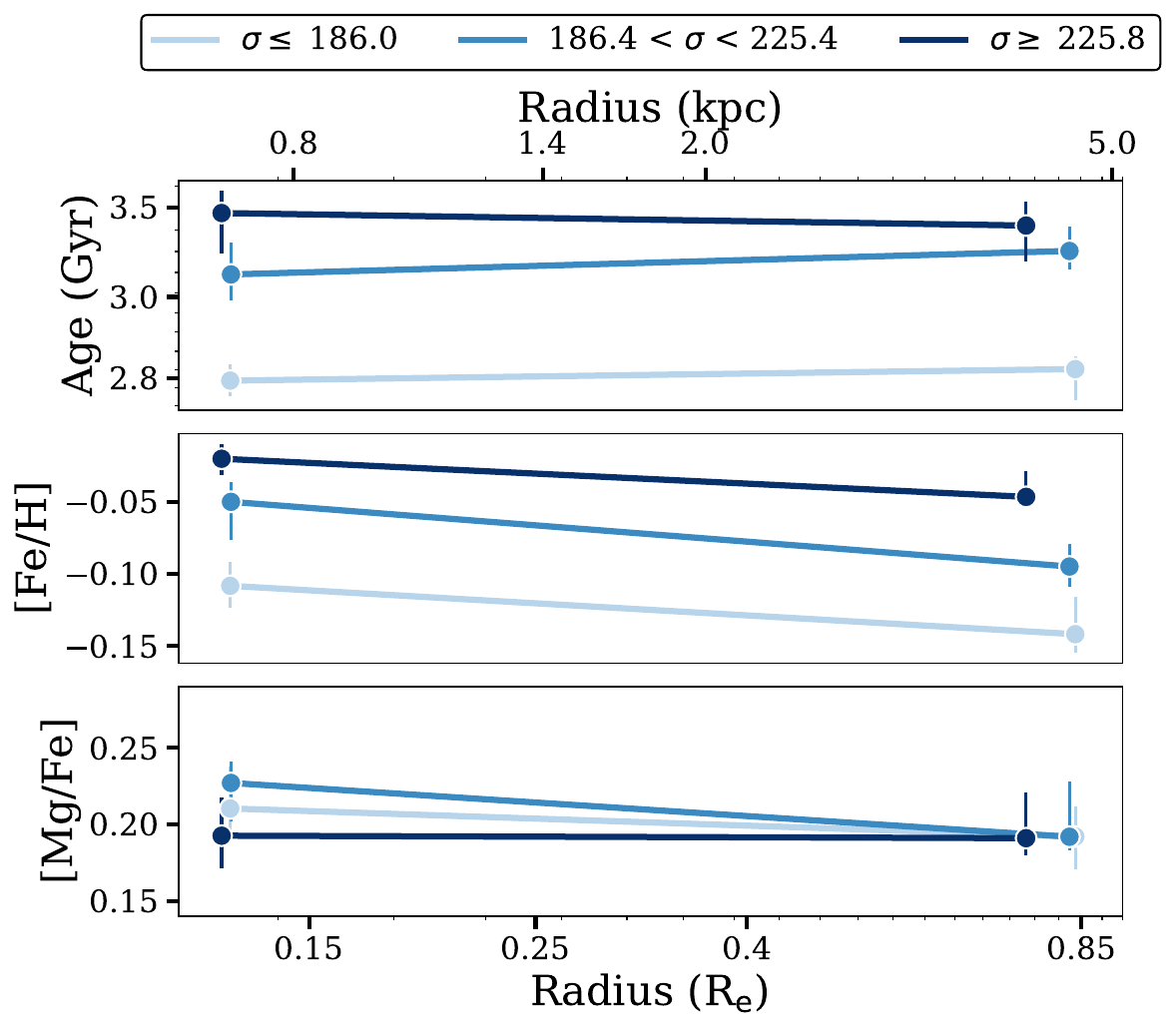}
    \caption{Stellar population parameters plotted as a function of radius ($R_{\rm e}$ and kpc) in convolved space.  Ages are shown in the top panel, [Fe/H] is shown in the middle, and [Mg/Fe] is shown on the bottom.  We split our sample into three velocity dispersion ($\sigma$) bins, with $\sigma$ in units of km s$^{-1}$.  These map approximately onto bins of stellar mass.  We plot the median stellar population parameters with uncertainties derived by bootstrap resampling in each $\sigma$ bin.}
    \label{fig:grad_age_bin}
\end{figure}

Our main results are shown in Figs~\ref{fig:grad_hist} and \ref{fig:grad_age_bin}.  In Fig.~\ref{fig:grad_hist}, we show histograms of the slope of each measured gradient (in convolved space).  The median is indicated in blue with shaded regions indicating the boostrapped uncertainties.  The dashed black line indicates where the median would lie if there were no gradient.  On average, we find that massive quiescent galaxies at $0.6\lesssim z \lesssim 1.0$ have age gradients consistent with being flat (median $\Delta\log(\mathrm{Age (Gyr)})/\Delta \log(R_{\rm e, convolved})  = 0.007^{+0.002}_{-0.004}$), mildly negative metallicity gradients (median $\Delta$[Fe/H] $/\Delta \log(R_{\rm e, convolved}) = -0.048^{+0.004}_{-0.009}$), and [Mg/Fe] gradients consistent with being flat (median $\Delta$[Mg/Fe] $/\Delta \log(R_{\rm e, convolved}) = -0.008\pm0.007$).  Our results are qualitatively consistent with Fig.~\ref{fig:paper_spectra}, where we can see that the H$\beta$ feature (sensitive to age) is visually not significantly different between the core and the outskirts.  However, the Fe feature near $5200$ \AA\ is deeper in the core than in the outskirts.

In Fig.~\ref{fig:grad_age_bin}, we show the values of age, [Fe/H], and [Mg/Fe] as a function of radius in units of $R_{\rm e}$ (in convolved space, see Section~\ref{sec:convolved_re}).  The values of each parameter have been split into equally sized stellar velocity dispersion ($\sigma$) bins for the age and [Fe/H] gradients (152 galaxies in each bin,\footnote{For the [Mg/Fe] gradients there are 72, 57, and 42 galaxies in each increasing $\sigma$ bin, respectively.} and where $\sigma$ is measured by the LEGA-C collaboration).  These map approximately onto bins of stellar mass.  The circles indicate the median in each bin with uncertainties derived by bootstrap resampling.  These are plotted at the median $R_{\rm e}$ in convolved space in each bin.  The data points are joined by lines to guide the eye.  From this figure, we see that our results of flat age gradients, negative [Fe/H] gradients, and flat [Mg/Fe] gradients hold across a wide range of $\sigma$.  We additionally recover the trends found in \cite{Beverage_2023} for quiescent galaxies in LEGA-C.  Specifically, they showed that age and [Fe/H] increase with $\sigma$, while [Mg/Fe] does not correlate with $\sigma$.

\begin{figure*}
    \centering
    \includegraphics[width=\textwidth]{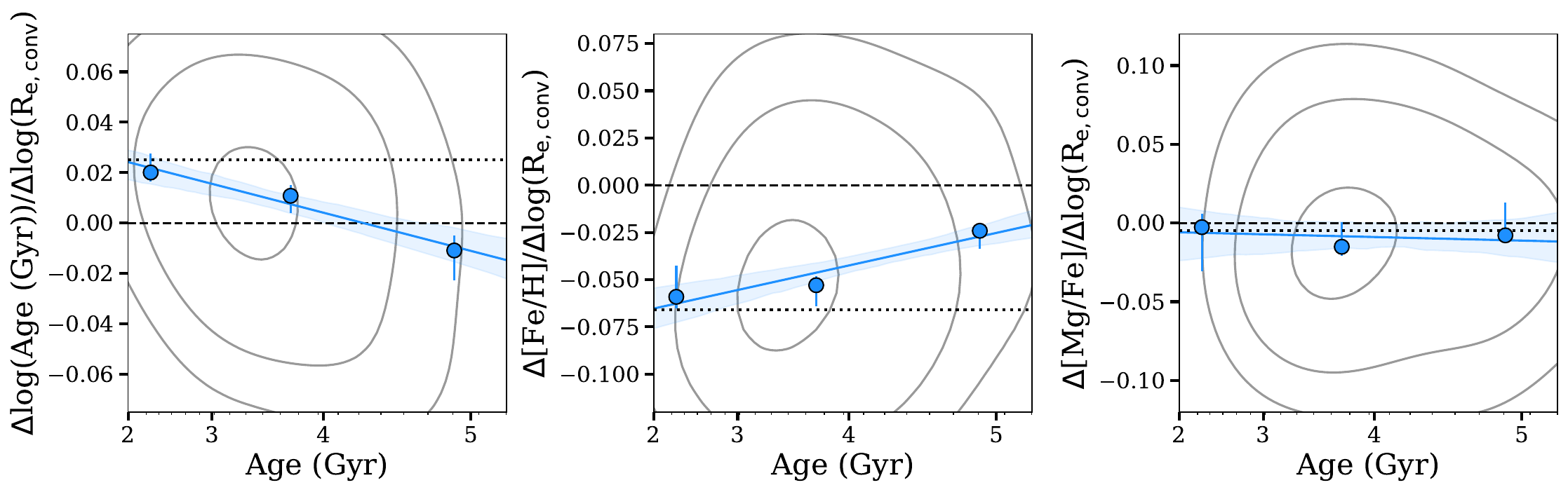}
    \caption{Spatially resolved stellar population gradients as a function of integrated age.  In each panel, the 50th, 68th, and 95th percentiles of the whole sample are represented by the contours.  The dashed line indicates where a flat gradient would lie.  The points show the median gradients in three evenly sized bins of integrated age, with uncertainties derived by bootstrap resampling.  The solid lines show linear fits to these running medians, and the shaded regions denote the $1\sigma$ uncertainties on these fits.  The dotted line indicates where no trend lies.}
    \label{fig:grad_vs_params}
\end{figure*}

In Fig.~\ref{fig:grad_vs_params}, we also examine our gradients as a function of integrated galaxy stellar age, by splitting the sample into three equally sized age bins (152 galaxies in each bin\footnote{For the [Mg/Fe] gradients there are 38, 57, and 76 galaxies in each increasing age bin, respectively.}).  We plot the running median in each age bin and the corresponding best linear fit to the running medians.  We also show our individual measurements as grey contours.  Splitting our sample into different age bins reveals intriguing trends.  In particular, we find that the age gradients are slightly positive for the youngest galaxies and flatten with stellar population age.  The metallicity gradients become weaker with increasing age, but are still negative at the oldest ages.  We find no trend between the [Mg/Fe] gradients and stellar population age (i.e. they are flat for all ages).  We comment further on these trends in Section~\ref{sec:discussion}.

Our findings are consistent with the predictions of \cite{Wu_2020}, who used LEGA-C to propose that positive age gradients were required in younger quiescent galaxies to explain the increase in galaxy size with age.  They are also consistent with the measurements of age gradients in young quiescent galaxies of \cite{D'Eugenio_2020} at similar $z$ and \cite{Pracy_2013} and \cite{Wu_2021} at low $z$ (and see also \citealt{French_2021}).  In particular, \cite{D'Eugenio_2020} measured Lick index gradients in the LEGA-C spectra for 17 post-starburst\footnote{We emphasize that our galaxies are \textit{not} necessarily PSB galaxies but are simply \textit{younger} quiescent galaxies.  Classifying galaxies in our sample as PSB is beyond the scope of this work.} (PSB) galaxies as well as a control sample of typical quiescent galaxies.  They found positive age gradients in the younger PSB galaxies.  On the other hand, in their control sample of older quiescent galaxies, they found that the central stellar populations were older and more metal-rich than the outer stellar populations.  However, their conclusions about the control sample were fundamentally limited by the use of only a few spectral indices; in contrast, our full-spectrum fits enable us to break the age--metallicity degeneracy, even for older quiescent galaxies (See Section~\ref{sec:caveats} and Appendix~\ref{sec:appendix_mocks}).  While we find that the older galaxies in our sample have flat age gradients, these results are otherwise qualitatively similar to what we find in this work.

On the other hand, \cite{Setton_2020} concluded that young quiescent galaxies at these redshifts have flat age gradients, based on their finding of a flat gradient of the H$\delta$ spectral index.  None the less, our results do not necessarily disagree with those of \cite{Setton_2020}.  We also examine the equivalent width of the H$\delta$ index in our spatially resolved spectra and find that the difference between H$\delta$ in the core and the outskirts is not significant.  However, when we make use of the full spectral wavelength range instead of one spectral feature, we are able to recover a positive age gradient in our youngest age bin.  This demonstrates that full spectral coverage is required to unveil gradients.  Moreover, this may not be surprising as, in addition to being sensitive to age, H$\delta$ is also sensitive to star formation history \citep{Worthey_1997}.  Thus, it does not perfectly trace the age gradient.  

We also examine our measured gradients in relation to several other physical parameters.  We find no significant trends between our measured gradients and intrinsic $R_{\rm e}$, spectroscopic redshift, stellar velocity dispersion, stellar mass, or Sérsic index.  The fact that we do not find any significant trends with total stellar velocity dispersion or stellar mass contrasts with the results of \cite{Spolaor_2010}, \cite{Gonzalez_Delgado_2015}, \cite{Ferreras_2019}, \cite{Santucci_2020}, and \cite{Yoon_2023} for low-$z$ galaxies, where varying dependence between mass and different stellar population gradients has been found.  For example, \cite{Spolaor_2010} found a positive trend between metallicity gradient and mass but no strong dependence for age and [Mg/Fe].  Alternatively, \cite{Ferreras_2019} found a weak dependence of metallicity gradient on velocity dispersion and a strong negative dependence of [Mg/Fe] gradient on velocity dispersion.  On the other hand, \cite{Sanchez_Blazquez_2007}, \cite{Pastorello_2014}, \cite{Gonzalez_Delgado_2015}, and \cite{Greene_2015} found that metallicity gradients do not depend on mass at low $z$. Additionally, studies making use of cosmological simulations find, in particular, no dependence of metallicity gradient on $M_*$ \citep{Kobayashi_2004, Cook_2016}, except perhaps at very large radii (\citealt{Cook_2016}, which are not reached by our measurements).

\begin{figure*}
    \centering
    \includegraphics[width=\textwidth]{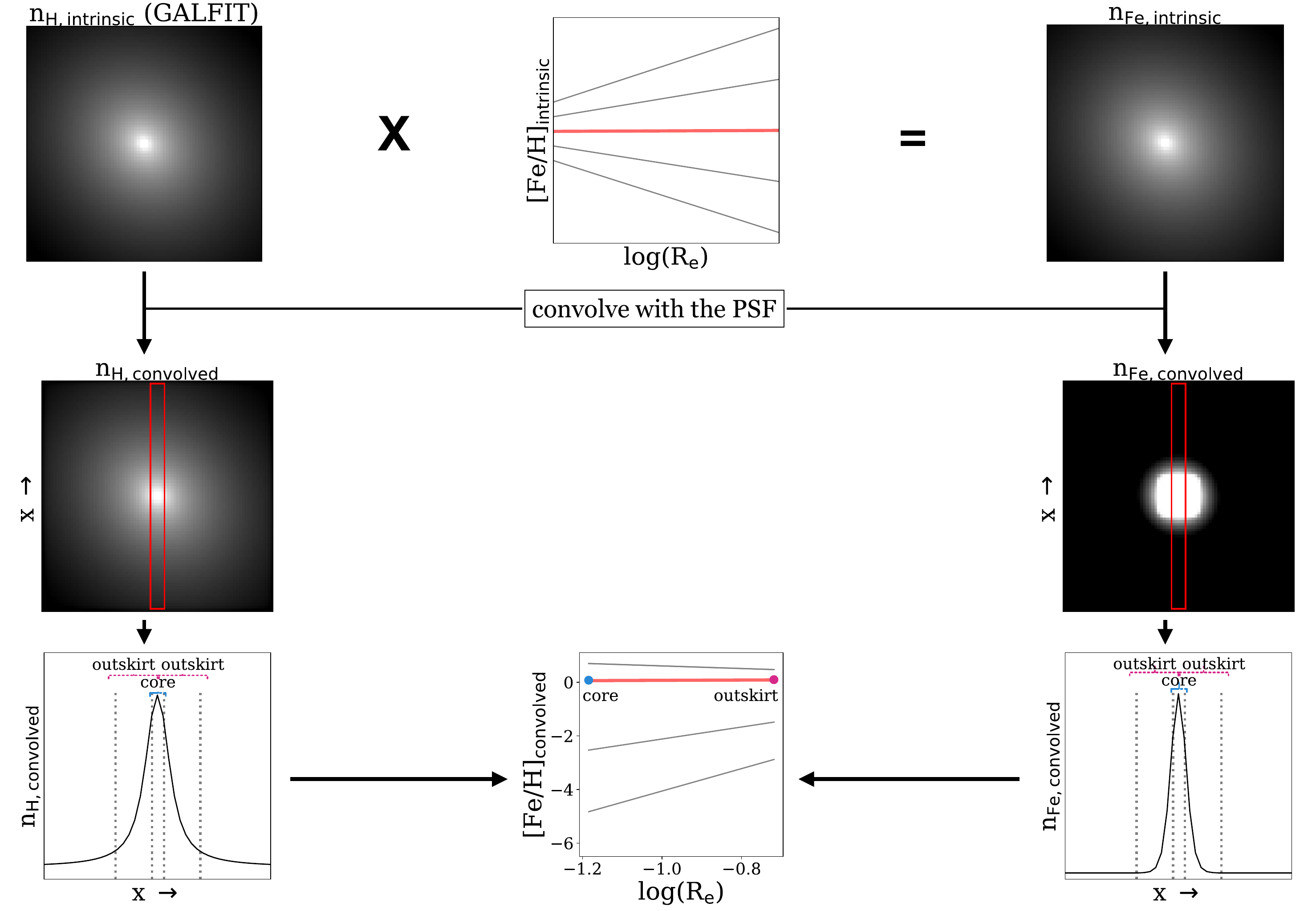}
    \caption{Graphical illustration of our intrinsic [Fe/H] gradient model.  For a detailed description of the procedure, see the main body of the text.  This process is repeated on a grid of 1000 intrinsic [Fe/H] gradient slopes.  The slope that gives the minimum absolute residual between our measured slope and the model convolved slope is then our predicted intrinsic gradient.}
    \label{fig:intrinsic_model_flowchart}
\end{figure*}

With this work, spatially resolved stellar population gradients have now been measured (with varying sample sizes) in massive quiescent galaxies out to $z\sim1$.  To obtain a detailed understanding of how these galaxies evolve from $z \sim 1$ to the present, we would need to quantitatively compare how the slopes of the stellar population gradients change over cosmic time.  However, comparing results between different studies is challenging as there are often significant differences in the definition of gradients, data quality, and methods to derive ages and metallicities.  In future work we will measure spatially resolved stellar population gradients using this same method out to $z \sim 3$ which will allow us to make a more homogeneous analysis of the evolution of gradients over cosmic time (see Section~\ref{sec:future_work}).  However overall, our current findings of typically negative metallicity gradients and flat age and [Mg/Fe] gradients for older galaxies are qualitatively consistent with results from the literature, both at low $z$ (e.g. \citealt{Mehlert_2003, Rose_2005, Kuntschner_2006, Sanchez_Blazquez_2007, Greene_2013, Greene_2015, Greene_2019, Pastorello_2014, Gonzalez_Delgado_2015, Martin_Navarro_2018, San_Roman_2018, Ferreras_2019, Oyarzun_2019, Zheng_2019, Lacerna_2020, Santucci_2020, Yoon_2023, Parikh_2024}) and at higher $z$ \citep{Jafariyazani_2020}.  This suggests that these stellar population gradients were in place by at least $z\sim1$ (with indications that this may even be true at $z\sim2$, \citealt{Jafariyazani_2020}) and are at least maintained until the current epoch.  

\subsection{Intrinsic metallicity gradients}\label{sec:intrinsic_gradients}
While our detection of negative metallicity gradients is significant, it is important to note that our observations are affected by the seeing of the LEGA-C observations.  To better understand the qualitative effect that the seeing has on our measurements, we develop a simple model to assess the metallicity gradients if we were to measure them in intrinsic space (i.e. in ideal conditions with no instrumental or atmospheric blurring).  

The model that we create here is similar in principle to the forward model in \cite{Suess_2019} and is illustrated in Fig.~\ref{fig:intrinsic_model_flowchart}.  We first create an `intrinsic 2D hydrogen profile'.  This is essentially the same \textsc{galfit} image that was generated in Section~\ref{sec:convolved_re} as here we have assumed that the hydrogen column density profile (i.e. $n_{\rm H}$) follows the light profile of the galaxy.

We use the hydrogen profile to create an intrinsic 2D [Fe/H] profile in linear space (i.e. $n_{\rm Fe}/n_{\rm H}$).  To do this, we assume that the $n_{\rm Fe}/n_{\rm H}$ profile follows the expression 
\begin{equation}\label{eq:intrinsic_grad}
    \log_{10}\left(\frac{n_{\rm Fe}}{n_{\rm H}}\right)_{\rm int} = m_{\rm int}\log(R_{\rm e}) + b_{\rm int} + \log_{10}\left(\frac{n_{\rm Fe}}{n_{\rm H}}\right)_\odot,
\end{equation}
where $m_{\rm int}$ and $b_{\rm int}$ are the intrinsic slope and $y$-intercept of the [Fe/H] gradient, respectively, and we obtain $\log_{10}(n_{\rm Fe}/n_{\rm H})_\odot = -4.5$ from \cite{Asplund_2009}.  We generate a grid of possible $m_{\rm int}$ (a grid of 1000 $m_{\rm int}$ with $-11 \lesssim m_{\rm int} \lesssim 11$) which spans a much wider range than our measured $n_{\rm Fe}/n_{\rm H}$ gradient slopes (since we expect the intrinsic gradients to be stronger than the convolved ones, see also \citealt{D'Eugenio_2020}).  We set $b_{\rm int}$\footnote{We note that, in this model, $b_{\rm int}$ is simply a scaling factor that indicates the metallicity at the centre of the galaxy.  It does not affect that slopes of the gradients, the key information in which we are interested here.} to the value of $n_{\rm Fe}/n_{\rm H}$ that we measure from the integrated fit.  We generate a 2D radius profile in units of pixels.  To avoid having a discontinuity in the centre of the simulated galaxy, we resample each pixel into $\sim 500$ subpixels, calculate the distance between each subpixel and the centre of the galaxy, and take the average of all of the subpixels in each pixel to be the distance between each pixel and the centre of the galaxy.  We plug the grid of $m_{\rm int}$'s (the parameter for which we are trying to fit), $b_{\rm int}$ (which we fix to the integrated value), and 2D $R_{\rm e}$ profiles (which we fix to the values we calculate here) into equation~\ref{eq:intrinsic_grad} to obtain a grid of intrinsic 2D $n_{\rm Fe}/n_{\rm H}$ profiles.

We multiply the intrinsic 2D $n_{\rm Fe}/n_{\rm H}$ profiles by the intrinsic 2D $n_{\rm H}$ profile to obtain intrinsic 2D $n_{\rm Fe}$ profiles.  We then convolve our intrinsic 2D $n_{\rm Fe}$ and $n_{\rm H}$ profiles with the best-fit seeing that we found in Section~\ref{sec:convolved_re} and we collapse these convolved images over the width of the VIMOS slit to obtain 1D profiles of $n_{\rm Fe}$ and $n_{\rm H}$.  

We weight the 1D profiles by the same weights that we use in our optimal extraction in Section~\ref{sec:extraction} and sum each profile.  We do this for each of our core (c) and outskirt (o) extractions to obtain a single value of each of $n_{\rm Fe, c}$, $n_{\rm H, c}$, $n_{\rm Fe, o}$, and $n_{\rm H, o}$ in convolved space for each grid point.  We calculate the model convolved slopes:
\begin{equation}
    m_{\rm conv} = \frac{\left(\log\left(\frac{n_{\rm Fe, o}}{n_{\rm H, o}}\right) - \log\left(\frac{n_{\mathrm{Fe,}\odot}}{n_{\mathrm{H,}\odot}}\right)\right) - \left(\log\left(\frac{n_{\rm Fe, c}}{n_{\rm H, c}}\right) - \log\left(\frac{n_{\mathrm{Fe,}\odot}}{n_{\mathrm{H,}\odot}}\right)\right)}{\log(R_{\rm e, o}) - \log(R_{\rm e, c})}
\end{equation}
and compare these to our measured slopes.  We take the grid point with the smallest absolute residual between our model $m_{\rm conv}$ and our observed slope to be the intrinsic [Fe/H] gradient of the galaxy.  

\begin{figure}
    \centering
    \includegraphics[width=\columnwidth]{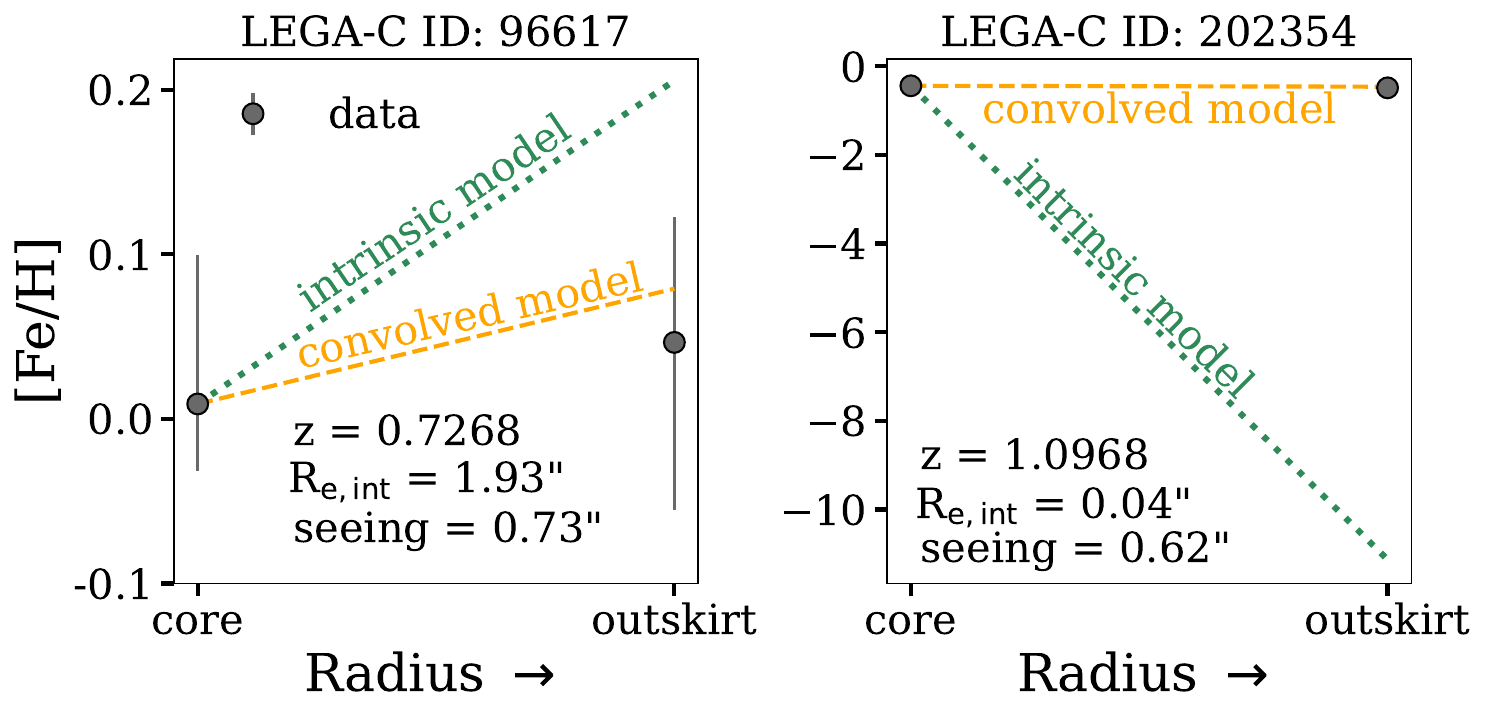}
    \caption{An illustration of how the instrinsic gradient model is applied to an example galaxy with a large $R_{\rm e}$ (top) and an example galaxy with a small $R_{\rm e}$ (bottom).  We emphasize that this is a qualitative illlustration of the intrinsic gradient slope that we obtain and the absolute [Fe/H] values are not meant to be taken at face value.  The grey circles indicate the spatially resolved [Fe/H] values that we measure from the LEGA-C spectra, the dashed line is the recovered convolved model [Fe/H] gradient, and the dotted line is the corresponding [Fe/H] gradient that we predict in intrinsic space.  In general, the intrinsic gradients are much steeper than those in convolved space.}
    \label{fig:intrinsic_model_example}
\end{figure}

\begin{figure}
    \centering
    \includegraphics[width=\columnwidth]{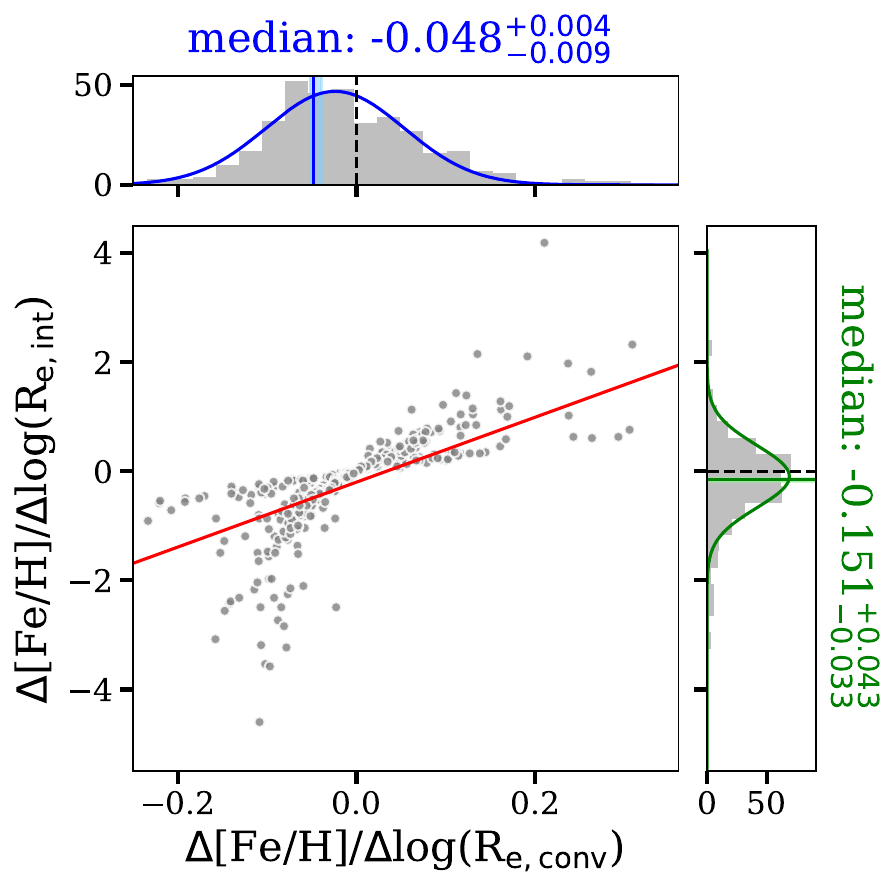}
    \caption{A comparison between our [Fe/H] gradients measured in convolved space to those that we model in intrinsic space.  The solid red line is the best fit from which we conclude that the intrinsic gradients are steeper than the observed ones by a factor of $5.96\pm0.31$.  A histogram of all intrinsic gradient predictions for our sample as well as a Gaussian fit to emphasize the shape of the distribution are shown on the right panel and the same for the measured gradients in convolved space is shown on top.  We show the median gradients as solid lines with shaded regions indicating uncertainties derived by bootstrap resampling.  The dashed lines show where flat gradients would lie.  We quote the median gradient values and bootstrapped uncertainties above each panel.  These are similar to the centre panel of Fig.~\ref{fig:grad_hist}.}
    \label{fig:intrinsic_model_compare}
\end{figure}

We stress that this model serves to qualitatively demonstrate the effect that the seeing may be having on our observations, and that these intrinsic gradients should not be taken as the true gradients for the individual galaxies.  As such, we do not model the uncertainties on our intrinsic gradient slopes.  Our method works well for $\sim 80$ per cent of our sample (i.e. the model does not recover the observed [Fe/H] gradient for 90 galaxies, due to a combination of their sizes being too small or their seeing being too large).  We show a qualitative illustration of how the model is applied to two example galaxies (one with a large $R_{\rm e}$ and one with a small $R_{\rm e}$ in Fig.~\ref{fig:intrinsic_model_example}).  We emphasize that this is a demonstration of how the best-fit convolved and intrinsic models look compared to the data.  We do not make use of the results for individual galaxies here and consider only the average trends. 

In Fig.~\ref{fig:intrinsic_model_compare}, we compare our observed [Fe/H] gradients to our predicted [Fe/H] intrinsic gradients.  We show histograms for each type of gradient on the top and to the right, similar to Fig.~\ref{fig:grad_hist}.  In the main panel, we fit a line to the comparison of the two types of gradients.  The slope of this line is the average factor by which the intrinsic [Fe/H] gradients are steeper than our observed gradients.  In general, we find that the intrinsic [Fe/H] gradients are steeper than our observed gradients, by a factor of $5.96\pm0.31$.  This indicates that the gradients we have measured here in massive quiescent galaxies at $0.6\lesssim z \lesssim 1.0$ are significant. 

We note that we only estimate the metallicity gradients here.  Developing a similar model for the spatially resolved ages would be much more complex.  However, we would expect any age gradients that we measure to be steeper in intrinsic space, with the effect being similar to what we find for the metallicity gradients.  In particular, positive age gradients would be more positive in intrinsic space and negative age gradients would be more negative.  Thus, we would expect our discussion in Section~\ref{sec:discussion} and our general conclusions to hold in intrinsic space. 

The `intrinsic gradients' that we measure here serve as a general demonstration of how much our measurements are affected by observational effects.  Nonetheless, it may still be interesting to compare our median intrinsic [Fe/H] gradient of $\Delta$[Fe/H] $/\Delta \log(R_{\rm e, intrinsic}) = -0.151^{+0.043}_{-0.033}$ to [Fe/H] gradients in the literature.  Again, we stress that it is difficult to compare results from different studies.  If we take the measurements at face value, however, we see that our intrinsic [Fe/H] gradients are generally shallower than [Fe/H] gradients at lower $z$ (e.g., \citealt{Greene_2015, Greene_2019}, etc.).  For example, \cite{Greene_2019} defined their [Fe/H] gradients in a similar way and found a median of $\Delta$[Fe/H]$/\Delta\log(R) = -0.26$.  We discuss this further in Section~\ref{sec:assembly}.

\section{Discussion}\label{sec:discussion}
\begin{figure*}
    \centering
    \includegraphics[width=\textwidth]{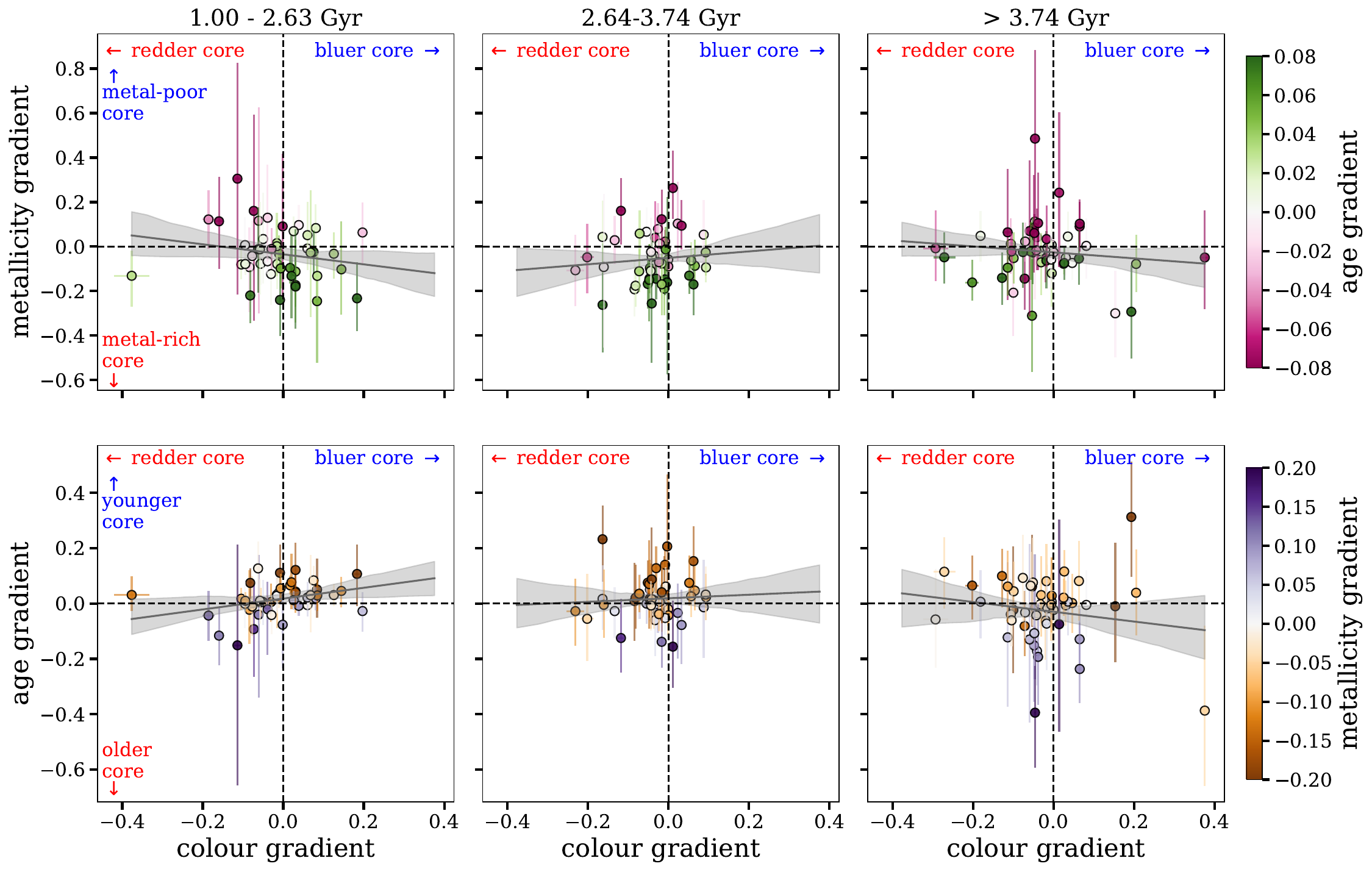}
    \caption{The relationship between our measured age and metallicity gradients and colour gradients.  On the $x$-axis of each panel, we use the ratio between $R_{\rm e}$ measured in HST/WFC3-\textit{F160W} images \citep{Cutler_2021} to $R_{\rm e}$ measured in HST/ACS-\textit{F814W} images as a proxy for colour gradients ($\log(R_{\mathrm{e,\textit{F160W}}}/R_{\mathrm{e,\textit{F814W}}})$).  In the top row, we show our [Fe/H] gradients on the $y$-axis ($\Delta{\rm [Fe/H]}/\Delta\log(R_{\rm e})$) and the symbols are colour coded by our age gradients ($\Delta \log({\rm Age (Gyr)})/\Delta\log(R_{\rm e})$).  In the bottom row, we show our age gradients on the $y$-axis and the symbols are colour coded by metallicity gradients.  We split this figure into three different age bins (the same as in Fig.~\ref{fig:grad_vs_params}).  We show a linear least-squares fit to the points in each panel with the shaded region showing the uncertainties on the fit.  We expect to see a positive trend if a particular gradient is the primary driver of the colour gradients.  We see that there is a positive (negative) trend between age ([Fe/H]) and colour gradient at the youngest ages.}
    \label{fig:cgrad_fig}
\end{figure*}

\subsection{Colour gradients}\label{sec:colour_grads}
The spatially resolved spectroscopic measurements that we perform here are intimately linked with spatially resolved measurements from photometry.  In cases where it has not been possible to obtain the high-quality spectra needed to resolve stellar populations, photometric data have been and are still used to examine spatially resolved stellar population properties.  

Differences in stellar populations across the spatial distribution of a galaxy can be observed photometrically as colour gradients.  In the local Universe, many studies have observed negative radial colour gradients in quiescent galaxies.  These gradients have largely been attributed to negative metallicity gradients (i.e. \citealt{Kormendy_1989} and references therein, \citealt{Peletier_1989, Franx_1990, Peletier_1990a, Davies_1993, Vazdekis_1997, Saglia_2000, La_Barbera_2005, Tortora_2010, Tortora_2011}), perhaps with a smaller contribution from age gradients (i.e. \citealt{La_Barbera_2009, Tortora_2010}).  

At higher redshifts, where even photometric observations have historically been more difficult to achieve, negative colour gradients have also been found on average in quiescent galaxies over a wide range in mass (i.e. \citealt{Wuyts_2010, Guo_2011, Szomoru_2012, Chan_2016, Liu_2017, Mosleh_2017, Suess_2019, Miller_2023a, Miller_2023b, Setton_2024}).  More specifically, it has been found that colour gradient strength increases with galaxy age, with younger quiescent galaxies having flatter colour gradients and older galaxies having more strongly negative colour gradients \citep{Suess_2020}.  This trend is further supported by the finding that PSB galaxies have flat colour gradients \citep{Roche_2010, Suess_2021}.  However, using colours alone, it is impossible to break the age--metallicity degeneracy \citep{Worthey_1994, Bruzual_2003, Gallazzi_2005}.  Thus, it is not necessarily clear what might be the origin of these colour gradients (or lack thereof) at higher redshifts.  In particular, a redder colour might be driven by an older, more metal-rich, or dustier stellar population.  For example, \cite{Miller_2023b} argued that metallicity gradients may be the primary driver due the fact that they find negative colour gradients similar to low-$z$ studies that have identified metallicity gradients as a primary driver, but this has yet to be shown definitively.  Alternatively, \cite{Miller_2023a} and \cite{Setton_2024} examined quiescent galaxies at $z\sim2$ and $\sim4$, respectively with photometry and low-resolution spectroscopy, and found indications that the negative colour gradients are driven by dust.  However, only a handful of quiescent galaxies were examined and metallicity was not considered.  Furthermore, with low resolution spectroscopy, it is extremely challenging to break the age--metallicity degeneracy.  

Our measurements can provide insight into the driver of observed colour gradients, at least at $0.6 \lesssim z \lesssim 1.0$.  We examine the comparison between our spectroscopic measurements and photometric colour gradients in Fig.~\ref{fig:cgrad_fig}.  In each panel, we show the ratio of $R_{\rm e}$ measured in \textit{HST}/WFC3-\textit{F160W} images \citep{Mowla_2019} to $R_{\rm e}$ measured in \textit{HST}/ACS-\textit{F814W} images \citep{Scoville_2007} on the $x$-axis.  This ratio is a proxy for colour gradient strength, as a positive colour gradient (i.e. bluer centre), for example, would result in smaller sizes at bluer wavelengths (i.e. F814W).  The \textit{F160W} sizes are obtained from \cite{Cutler_2021} and \textit{F814W} sizes are measured by the LEGA-C collaboration, both using \textsc{galfit} \citep{galfit}.  We exclude \textit{F160W} sizes flagged as `bad' in \textsc{galfit}.  Unfortunately, the majority of the \textit{F160W} sizes for our sample are not well-fitted by \textsc{galfit} so we are only able to calculate this ratio for a fraction of our sample.  This is due to a combination of the imaging from \cite{Cutler_2021} being relatively shallow and the fact that they do not have imaging for the entire UltraVISTA \citep{McCracken_2012} catalogue.  In the top row, we show our measured [Fe/H] gradients on the $y$-axis and the symbols are colour coded by our measured age gradients.  In the bottom row, we show the reverse -- measured age gradients are on the $y$-axis and the symbols are colour coded by measured [Fe/H] gradients.  We have split the sample into three age bins (the same ones as in Fig.~\ref{fig:grad_vs_params}).  In each panel, we show a linear least-squares fit to the points, with the shaded region showing the uncertainties on the fit.  If a particular gradient is the primary driver of the colour gradient, then we expect to see a positive trend between that gradient and the colour gradient strength.

For the younger galaxies ($1$ Gyr $\leq$ age $< 2.63$ Gyr, left panel of Fig.~\ref{fig:cgrad_fig}), we find a positive trend between age and colour gradients and a negative trend between [Fe/H] and colour gradients.  This suggests that the flat colour gradients found in young quiescent galaxies by \cite{Suess_2020} may actually be the result of the age and metal gradients compensating each other.  In particular, a positive age gradient is expected to contribute to a bluer core and redder outskirts while a negative metallicity gradient results in the opposite.  Thus, in colour space, the two gradients effectively cancel each other out.  We note that it is beyond the scope of this paper to quantify how positive an age gradient would have to be to compensate a negative metallicity gradient, however our results are qualitatively indicative of this scenario.  

For the older galaxies ($> 2.63$ Gyr, middle and right panels of Fig.~\ref{fig:cgrad_fig}), we see a flat relationship between colour and age gradients.  The relationship with [Fe/H] gradient is also mostly flat.  We cannot conclude much from this figure, which may not be surprising as the observed age and metal gradients are weaker for these galaxies and the uncertainties on the colour gradients are large.  However, while no clear trend is seen with [Fe/H] gradients, the lack of trend between age and colour gradients may suggest that colour gradients at older ages are most likely driven primarily by [Fe/H] gradients.  Of course, though our observed stellar population gradients can, qualitatively, explain the observed colour gradients, we cannot rule out that dust is contributing to the colour gradients as well (see \citealt{Miller_2023a, Setton_2024}) as we do not consider the effects of dust in our analysis.  In particular, if the cores of these galaxies are dustier than their outskirts, then their observed colour gradients may be driven primarily by dust instead of age or metallicity.  More work is needed to determine the primary driver of colour gradients at $z > 0$.

\subsection{Implications for galaxy evolution}\label{sec:assembly}
In our analysis, we find that there are distinct differences in the stellar population gradients of galaxies at different ages, with younger galaxies having positive age gradients and stronger metallicity gradients, and older galaxies having flat age gradients and weaker metallicity gradients.  This suggests that the cores of the younger galaxies are younger and more metal-rich than their outskirts, and the cores of older galaxies, while still being more metal-rich, are the same age as their outskirts.  Furthermore, we find flat [Mg/Fe] gradients at all ages, which implies that the core and outer regions of the galaxies in our sample have similar star formation time-scales (see e.g. \citealt{Maiolino_2019}).  However, we caution that it is difficult to interpret our [Mg/Fe] gradients as we measure them for a much smaller sample compared to our age and [Fe/H] gradients.  We note that there is a significant amount of scatter in our gradient measurements for individual galaxies, however these average trends still have interesting implications for galaxy assembly histories.  Here we will discuss the possible scenarios suggested by our results, first for the younger galaxies and then for the older galaxies. 

\subsubsection{Young quiescent galaxies}\label{sec:young}
We find that massive quiescent galaxies in the youngest age (i.e. $1$ Gyr $\leq$ age $< 2.63$ Gyr old) bin of our sample tend to have positive age gradients and negative metallicity gradients.  In other words, their cores tend to be younger and more metal-rich than their outskirts.  Thus, this core stellar population may be indicative of a central starburst that occurred just before the galaxy stopped forming stars.  These observed age gradients were predicted observationally by \cite{Wu_2020} for young, compact quiescent galaxies, to explain their observed increase in half-light radii with age.  As the central starburst fades, the older, more extended population will become more dominant, resulting in an increase of the half-light radius (see also \citealt{Whitaker_2012, Yano_2016, Almaini_2017, Maltby_2018}).  \cite{Pathak_2021} also showed that the majority of their IllustrisTNG-simulated young quiescent galaxies have positive age gradients as a result of recent central starbursts (although a significant portion of their young quiescent galaxy sample was also consistent with having flat age profiles).  In this context, it is furthermore interesting to note that positive age gradients may also be present in the compact star-forming population, as their star formation is more centrally concentrated than the stellar population that is already present (e.g., \citealt{Barro_2016b, Barro_2017, Tadaki_2017, Popping_2017}).  Therefore, compact star-forming galaxies which were rapidly quenched may indeed be the progenitors of young quiescent galaxies \citep{Barro_2013, Barro_2014a, Barro_2014b, Barro_2016a, Barro_2017, D'Eugenio_2020}.

Our results provide insights into the quenching mechanisms of young quiescent galaxies.  In particular, we can discriminate between two popular suggested mechanisms in the literature at these redshifts: wet compaction events or wet, gas-rich major mergers.  In the case of wet compaction, clumps of gas migrate towards the centre, compacting the galaxy by decreasing the $R_{\rm e}$.  This occurs preferentially at high redshifts.  Galaxies may then experience extended star formation in the outskirts after the gas in the core is depleted.  In particular, simulations have shown that following the compaction phase, compact galaxies may be encircled by an extended disc of young stars and low-density gas.  Quenching post-compaction is then expected to occur inside--out \citep{Bell_2008, Bell_2012, Fang_2013, Li_2015, Tacchella_2015, Tacchella_2016, Zolotov_2015, Gonzalez_Delgado_2016, Lin_2017, Ellison_2018, Sanchez_2018, Spilker_2019, Breda_2020, Nelson_2021, Abdurrouf_2023, Avila_Reese_2023, Lapiner_2023}.  In this way, a compaction event would be expected to produce negative age gradients \citep{Dekel_Burkert_2014, Zolotov_2015, Tacchella_2016}.  On the other hand, mergers can trigger a strong starburst by driving gas to the centre of the galaxy, producing a positive age gradient \citep{Hopkins_2008, Snyder_2011, Wellons_2015, Pathak_2021}.  Thus, our results support a merger-triggered central starburst event, or another quenching mechanism that results in a central starburst just prior to quenching.  In this regard, our conclusions are consistent with those of \cite{Setton_2020}.

\subsubsection{Growth of the quiescent galaxy population}\label{sec:old} 
In contrast to the mildly positive age gradients that we find in younger galaxies, older quiescent galaxies in our sample tend to have flat age and [Mg/Fe] gradients and weaker (but still negative) [Fe/H] gradients.  In this section, we will discuss what these observations may imply for the evolution of individual quiescent galaxies, as well as for the growth of the quiescent galaxy population over time.

Our finding that the outskirts of massive quiescent galaxies have lower metallicities compared to their cores is supportive of inside--out formation, which has been suggested by many previous studies (e.g., \citealt{van_der_wel_2008, Bezanson_2009, Hopkins_2009, Naab_2009, Oser_2010, van_Dokkum_2010, Moster_2013, van_de_Sande_2013, Abramson_2014, Somerville_2015, Nelson_2016, Hill_2017, Tacchella_2019, Suess_2019, Suess_2020, Santucci_2020, Pulsoni_2021, Conselice_2022, Beverage_2023}).  For example, the measurement of small sizes of distant quiescent galaxies points to a scenario in which quiescent galaxies build up their outskirts by the accretion of lower-metallicity, younger galaxies over cosmic time (e.g., \citealt{Trujillo_2004, Trujillo_2006, Franx_2008, van_der_wel_2008, van_der_Wel_2014, Bezanson_2009, Naab_2009, van_Dokkum_2010}).  In this scenario, the lack of an age gradient in older quiescent galaxies may be explained by central rejuvenation, as discussed in Section~\ref{sec:young}.  This may lower the average age (and [Mg/Fe]) in galaxy cores to a similar age (and [Mg/Fe]) to that of the accreted galaxies.  Thus, our observed gradients may support inside--out growth driven by \textit{minor mergers} (e.g., \citealt{Bezanson_2009, Naab_2009, Oser_2010}, etc.) associated with central rejuvenation.  The minor merger mechanism is also supported by our finding of shallower intrinsic [Fe/H] gradients compared to low-$z$ studies such as \cite{Greene_2019} (Section~\ref{sec:intrinsic_gradients}), as minor mergers would cause [Fe/H] gradients to become steeper towards lower $z$ \citep{Naab_2009, Oser_2010}.  On the other hand, our results disfavour inside--out growth driven by \textit{late-time star formation} (i.e., \citealt{Abramson_2014, Nelson_2016}).  In this case, the outskirts of quiescent galaxies would be built up by this late-time star formation, resulting in younger stellar populations in the outskirts (i.e. a negative age gradient). 

Another popular scenario that can explain the size growth of the quiescent galaxy population over cosmic time is that of progenitor bias, where galaxies that quench at later times are larger \citep{van_Dokkum_2001, Carollo_2013, Poggianti_2013}.  Age and metallicity gradients are generally expected in this case \citep{Larson_1974, White_1980, Kobayashi_2004}, and the models of \cite{Pipino_2008} and \cite{Pipino_2010} have shown that it is possible to produce negative [Fe/H] gradients and flat [$\alpha$/Fe] gradients.  Thus, in this scenario, individual galaxies evolve passively and the gradients are directly inherited from the star-forming phase.  Additionally, galaxies in our older age bin quenched at earlier times ($z_{\rm form} \gtrsim 2$), while galaxies in our younger age bins quenched at later times ($0.9 \lesssim z_{\rm form} \lesssim 1.4$).  High-$z$ star-forming galaxies have indeed been found to have negative metallicity gradients (e.g., \citealt{Jones_2010, Jones_2012, Jones_2015}) and negative or flat age gradients \citep{Tripodi_2024, Shen_2024}, consistent with our findings.  Furthermore, colour gradients of star-forming galaxies have been found to become stronger at lower $z$ \citep{Mosleh_2017, Suess_2019, Suess_2019b}.  Thus, our findings are qualitatively consistent with the progenitor bias scenario as well. 

In summary, our observed gradients are consistent with both the minor merger and progenitor bias scenarios.  In fact, it is likely that both are playing a role, as suggested by, for example, \cite{van_der_wel_2008}, \cite{Hopkins_2010}, \cite{Valentinuzzi_2010}, \cite{Oser_2012}, \cite{Newman_2012}, \cite{Nipoti_2012}, and \cite{Barro_2013}.  For instance, it may be that mergers are only important for the most massive galaxies ($\log(M_*/M_\odot) > 11$, \citealt{Carollo_1993, Rodriguez_Gomez_2016}).  It is also possible that different mechanisms can contribute to the assembly of specific kinds of galaxies at specific redshifts \citep{Belli_2014, Belli_2015, Belli_2017, Wellons_2015, Wellons_2016}.  For example, \cite{Suess_2021} found progenitor bias to be primarily important in slow-quenching green valley galaxies at low redshifts.  None the less, \cite{HM_Beverage} and \cite{Kriek_2024} also argue for some contribution from major mergers to explain the metal evolution and dynamical masses of distant quiescent galaxies.

The scatter in our gradient measurements could be used as an additional probe to understand the importance of the different processes.  However, the uncertainties in our individual measurements are quite large, and the significant amount of scatter in our observed gradients can bee attributed primarily to these measurement uncertainties.  Thus, further observations are required to conclude which process may be dominant. 
  
\subsubsection{Future work}\label{sec:future_work}
Our findings contribute to a vast body of work that aims to understand the build-up of the quiescent galaxy population.  However, much still remains to be done to understand which of the scenarios that we have outlined in the previous sections may describe the primary mechanism of galaxy assembly, and to learn how the negative [Fe/H] gradients originate.  There are two main ways in which we can expand upon this work: going to higher redshifts and increasing our spatial resolution.

Examining spatially resolved stellar population gradients at higher redshifts will allow us to understand how these gradients have evolved over time.  For example, \cite{Suess_2019b} found that colour gradients at $z \lesssim 1.0$ (i.e. the regime explored in this study) remain relatively constant, consistent with the lack of trend that we see between our observed gradients and $z$ (Section~\ref{sec:observed_gradients}).  However, at higher redshifts ($1.0\lesssim z \lesssim 2.5$), they found that colour gradients evolve rapidly.  It is important to investigate whether similar evolution happens in spectroscopic gradients.  Understanding how gradients change over time will in turn allow us to better constrain the primary mechanism of galaxy assembly.  In particular, in the scenario of hierarchical formation, it is predicted that the compact central regions of galaxies are formed at $z\sim2-3$ and are built up by accretion at later times \citep{Oser_2010, Rodriguez_Gomez_2016}.  In this case, we would expect minor mergers to strengthen [Fe/H] gradients toward $z\sim0$.  However, for example, \cite{Jafariyazani_2020} found a steeper [Fe/H] gradient in an individual lensed galaxy at $z\sim2$ than measurements at low $z$ and our predicted intrinsic [Fe/H] gradients (Section~\ref{sec:intrinsic_gradients}).  Probing closer to the epoch of formation of these compact cores with a larger sample may allow us to point to a dominant assembly mechanism.  In an upcoming study, we will extend the redshift range of this analysis up to $z\sim3$ with ultra-deep, medium resolution spectra from the Spectroscopic Ultradeep Survey Probing Extragalactic Near-infrared Stellar Emission (SUSPENSE), a Cycle 1 \textit{JWST}/NIRSpec survey of 20 distant quiescent galaxies at $1 < z < 3$ \citep{Slob_2024}.  

Increased spatial resolution is also important for the accurate interpretation of stellar population gradients.  In particular, \cite{Oyarzun_2019} found that average radial metallicity profiles of low-$z$ galaxies from the Mapping Nearby Galaxies at APO (MaNGA) survey are not linear and thus a single gradient value does not fully characterize metallicity profiles.  Specifically, they found that metallicity profiles flatten in galaxy outskirts.  This points to stellar accretion from merging galaxies, showing that higher resolution data can also provide key insights about galaxy assembly mechanisms.  In this study we do not recover detailed stellar population profiles (i.e. with more than two resolution elements).  This is primarily because we do not have the required S/N in smaller spatial bins to reliably recover ages and metallicities.  Higher S/N, higher spatial resolution spectra can make this possible.  We hope to achieve this in future work with \textit{JWST}-SUSPENSE as well.  Additionally, IFU data from the Multi Unit Spectroscopic Explorer (MUSE) or the Enhanced Resolution Imager and Spectrograph (ERIS) on the \textit{VLT} may provide more detailed measurements. 

\subsection{Caveats}\label{sec:caveats}
The depth and resolution of the LEGA-C spectra in combination with the \textsc{alf} full spectrum models have allowed us to measure statistically significant spectroscopic age and metal gradients beyond the low-$z$ Universe for the first time.  However, there are several caveats that we address here and that should be kept in mind when interpreting our results.  

The most impactful consideration is the fact that our data are highly blurred by atmospheric seeing and instrumental effects, combined with the fact that the VIMOS slit and our 1D spectral extraction regions are relatively narrow compared to the size of the galaxy.  Thus, contamination between the core and outskirt regions is a concern for our gradient measurements.  However, we would expect any gradient that we are able to detect in the convolved space of the LEGA-C data to be even stronger in intrinsic space.  Therefore, the fact that we are able to detect differences between the central and outer regions of the galaxies at all is extremely encouraging and indicates that the gradients are significant and likely stronger than what we report here.  This fact and the effect of any potential contamination between the core and the outskirts is also demonstrated by our intrinsic model in Section~\ref{sec:intrinsic_gradients}.  Future work with data from the space-based \textit{JWST} will partially mitigate this issue.

Additionally, we argue that our optimal extraction procedure in Section~\ref{sec:extraction} does not affect our conclusions.  Recall from Section~\ref{sec:extraction} that we do not include the outermost wings of the spectra (the outermost 3 per cent of the spectral rows on each side of the profile, see shaded regions in the right panel of Fig.~\ref{fig:paper_extraction}), as we find that they introduce a lot of noise.  Our detailed extraction routine thus maximizes our sample size while also maintaining a similar S/N between the core and the outskirts for most objects.  We perform several tests and find that modifications to the exact extraction do not impact our conclusions.  For example, we test including all of the spectral rows, using the profile itself to weight the extraction, and testing different proportions of rows in the core and the outskirts  (i.e. while discarding the outer 6 per cent of the wings, we have tested a 47 per cent/47 per cent split and a 44 per cent/50 per cent split).  In each test, we are able to recover the same trends as presented in Section~\ref{sec:results}, with the median values for each gradient in each test within uncertainties of each other.  Thus, the wings of the spectra do not contribute additional information to the gradients in these galaxies with the S/N that we have and our precise extraction routine does not impact our conclusions. 

Another concern with this method of spectral binning is contamination of the core spectrum with outskirt information and vice versa, due to inclination effects.  For galaxies which we observe face-on, this is not an issue as the core and outskirts are distinct in the 2D spectra.  However, for galaxies which we observe edge-on or with some inclination, contamination may be an issue.  We find that the distribution of gradients in age, metallicity, and Mg are approximately the same in bins of increasing axis ratio, so we do not expect this to be a significant issue.  

We also perform tests to understand whether oversubtraction of the sky background could result in the false detection of gradients.  oversubtraction would preferentially affect the outskirts and artificially weaken the absorption lines compared to the core, potentially creating a gradient where none exists.  However, this would affect all features across the spectrum equally, and we see in the inset panels of Fig.~\ref{fig:paper_spectra} that this is not the case.  Specifically, the H$\beta$ feature is not significantly different between the core and the outskirts, whereas the comparably-deep Fe52 feature is distinct between the core and outskirt regions.  Moreover, we test whether we would be able to recover a gradient if the sky background was oversubtracted.  For a subsample of galaxies, we create a simulated oversubtracted outskirt spectrum by subtracting a constant value of 1 per cent of the maximum of the core spectrum from the core spectrum and fit this with \textsc{alf}.  In this test, we typically find flat age, [Fe/H], and [Mg/Fe] gradients.  Thus, our detection of gradients is not the result of oversubtraction of the sky background.   

Finally, the age--metallicity degeneracy is a well-known issue in stellar population studies, whereby the effects of age and metallicity on integrated light can be confused \citep{Worthey_1994, Bruzual_2003, Gallazzi_2005}.  This can severely bias measurements of ages and metallicities.  We confirm that the depth, resolution, and spectral coverage of the LEGA-C spectra allow us to disentangle the effects of age and metallicity on our spectra.  In particular, our gradient measurements are not impacted by the age--metallicity degeneracy.  We show this in Appendix~\ref{sec:appendix_mocks}, where we simulate different kinds of age and metallicity gradients in mock spectra, and find that we are able to recover the input gradients. 

\section{Summary and conclusion}\label{sec:conclusions}
In this paper, we presented spatially resolved stellar population gradients for 456 massive quiescent galaxies at $0.6 \lesssim z \lesssim 1.0$ from the LEGA-C survey.  We extracted 1D integrated spectra as well as 1D spectra comprising the core and outskirts of each galaxy from the LEGA-C 2D spectra.  We measured spatially resolved ages, [Fe/H], and [Mg/Fe] by fitting the spectra with a flexible, full-spectrum SPS model, \textsc{alf}.  We also forward-modelled what we would expect our [Fe/H] gradients to look like in unblurred, intrinsic space.  Our main conclusions are summarized here:
\begin{itemize}
    \item On average, massive quiescent galaxies at these redshifts have flat age (median $\Delta\log(\mathrm{Age (Gyr)})/\Delta \log(R_{\rm e, convolved})  = 0.007^{+0.002}_{-0.004}$) and [Mg/Fe] (median $\Delta$[Mg/Fe] $/\Delta \log(R_{\rm e, convolved}) = -0.008\pm0.007$) gradients and negative [Fe/H] gradients (median $\Delta$[Fe/H] $/\Delta \log(R_{\rm e, convolved}) = -0.048^{+0.004}_{-0.009}$), corresponding to more iron-rich galaxy cores.  These results are consistent with what has been found for quiescent galaxies in the literature, both in the local Universe (e.g., \citealt{Greene_2015, Martin_Navarro_2018}) and at $z\sim 2$ \citep{Jafariyazani_2020}.  
    \item In intrinsic space, we find that our [Fe/H] gradients are, on average, expected to be a factor of $6.0\pm0.3$ steeper than in observed space.  Thus, our intrinsic gradients are much steeper than our measured gradients, with a median of $\Delta$[Fe/H] $/\Delta \log(R_{\rm e, convolved}) = -0.15^{+0.04}_{-0.03}$.
    \item When we split our galaxy sample into three equal age bins, we find that the youngest quiescent galaxies have positive age gradients, while the older quiescent galaxies have flat age gradients.  The [Fe/H] gradients weaken with age, but remain negative at all ages.  The [Mg/Fe] gradients remain flat for all age bins, implying that the outer and inner regions formed over similar timescales (see e.g. \citealt{Maiolino_2019}).
    \item These findings suggest that photometrically measured flat colour gradients in young quiescent galaxies (and PSB galaxies) at high redshifts \citep{Suess_2020, Suess_2021} may be the result of the positive age and negative metallicity gradients compensating each other.  Meanwhile, negative colour gradients found in older quiescent galaxies \citep{Wuyts_2010, Guo_2011, Szomoru_2012, Chan_2016, Liu_2017, Mosleh_2017, Suess_2019, Suess_2020, Miller_2023a, Miller_2023b, Setton_2024} are likely driven primarily by metallicity gradients.
    \item Furthermore, our results indicate that the gradients we observe in young quiescent galaxies may be due to a recent central starburst.   These findings support a merger-triggered central starburst event as the primary quenching mechanism for young quiescent galaxies out to $z\sim1$ (also predicted to produce/maintain positive age gradients).  
    \item Our finding of negative metallicity gradients is supportive of a picture of inside--out growth via minor mergers of less massive, lower metallicity satellite galaxies.  On the other hand, our results are also consistent with the progenitor bias scenario, with gradients being inherited directly from the star-forming phase.  In reality, it is likely that both minor mergers and progenitor bias are playing a role in the assembly of massive quiescent galaxies.  
\end{itemize}

This work represents the first study of detailed spectroscopic stellar population gradients in a statistically significant quiescent galaxy sample beyond the low-redshift Universe.  These observations show that simply using colour gradients does not capture the full breadth of stellar population gradients in distant galaxies and have given us deeper insights into the possible mechanisms of galaxy formation and evolution.  However, we are still limited by observational effects due to ground-based observations and moderate redshifts, as the rest-frame optical shifts into the near-infrared beyond $z = 1$.  With its unprecedented sensitivity at near-infrared wavelengths and its high spatial resolution, \textit{JWST} will allow us to significantly improve upon the current work and push it to higher redshifts in the near future.  

\section*{Acknowledgements}
We thank the anonymous referee for a useful report that improved this manuscript.  We thank the LEGA-C team for making their dataset public.  We also thank Piyush Sharda, Jesse van de Sande, Ivana van Leeuwen, and Colin Yip for useful conversations.  This work was performed using the compute resources from the Academic Leiden Interdisciplinary Cluster Environment (ALICE) provided by Leiden University.  This work also used the Dutch national e-infrastructure with the support of the Samenwerkende Universitaire Rekenfaciliteiten (SURF) Cooperative using grant no. EINF-6344 which is financed by the Dutch Research Council (NWO). MK acknowledges funding from the NWO through the award of the Vici grant VI.C.222.047 (project 2010007169) and National Science Foundation (NSF) Astronomy and Astrophysics Research Grant (AAG) AST-1909942.  RB acknowledges support from the Research Corporation for Scientific Advancement (RCSA) Cottrell Scholar Award ID No: 27587 and from the National Science Foundation NSF-CAREER grant \# 2144314.  FDE acknowledges support by the Science and Technology Facilities Council (STFC), by the European Research Council (ERC) through Advanced Grant 695671 `QUENCH', and by the UK Research and Innovation (UKRI) Frontier Research grant RISEandFALL.  PEMP acknowledges support from the NWO through the Veni grant VI.Veni.222.364.  P.F.W. acknowledges funding through the National Science and Technology Council grant 111-2112- M-002-048-MY2.  A.G. acknowledges support from Italian National Institute for Astrophysics (INAF)-Minigrant-2022 `LEGA-C' 1.05.12.04.01.

%%%%%%%%%%%%%%%%%%%%%%%%%%%%%%%%%%%%%%%%%%%%%%%%%%
\section*{Data Availability}
This study makes use of data from the LEGA-C survey.  The two-dimensional and reduced one-dimensional spectra can be obtained from the ESO Science Archive Facility (\url{http://archive.eso.org/eso/eso_archive_main.html}).  The reduced spectra and catalogue have been released by ESO \newline (\url{http://archive.eso.org/cms/eso-archive-news/Third-and-final-release-of-the-Large-Early-Galaxy-Census\\-LEGA-C-Spectroscopic-Public-Survey-published.html}) and are also available here: \url{https://users.ugent.be/~avdrwel/research.html#legac}.  For more details, please see \cite{van_der_Wel_2016}, \cite{van_der_Wel_2021}, and \cite{Straatman_2018}.

Individual stellar population ages and elemental abundances will be published in an upcoming work.  Other data products generated in the course of this work (i.e. individual gradients and intrinsic model gradients) will be made available upon reasonable request. 



\bibliographystyle{mnras}
\bibliography{references} % if your bibtex file is called example.bib

%\begin{thebibliography}{99}
%\end{thebibliography}

%%%%%%%%%%%%%%%%%%%%%%%%%%%%%%%%%%%%%%%%%%%%%%%%%%

%%%%%%%%%%%%%%%%% APPENDICES %%%%%%%%%%%%%%%%%%%%%

\appendix

\section{Mock spectra tests}\label{sec:appendix_mocks}
\begin{figure*}
    \centering
    \includegraphics[width=\textwidth]{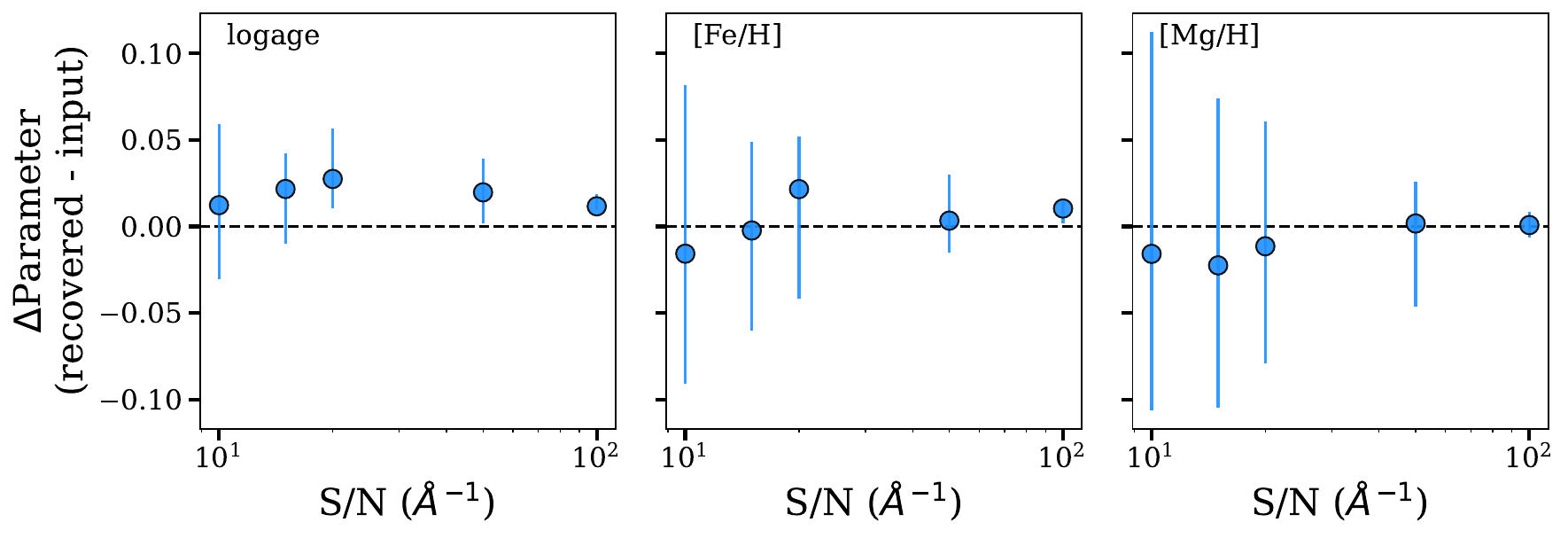}
    \caption{Mock spectra tests to determine the minimum S/N per \AA\ that we need in order to recover age (left panels), [Fe/H] (middle panels), and Mg abundances (right panels) when we fix the other elemental abundances to their input values.}
    \label{fig:SN_mocks}
\end{figure*}

\begin{figure}
    \centering
    \includegraphics[width=\columnwidth]{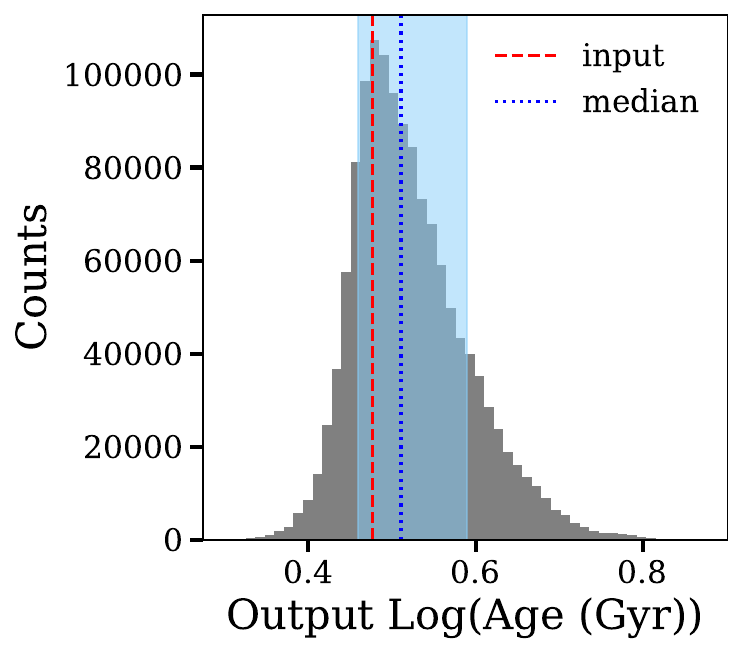}
    \caption{The distribution of the stacked MCMC chains for the recovered age from 40 mock spectra with S/N $ = 20$ \AA$^{-1}$.  The median of this distribution is indicated by the dotted line, with the shaded regions indicating the uncertainties.  The dashed line indicates the input age value.}
    \label{fig:mock_age_test}
\end{figure}

\begin{figure}
    \centering
    \includegraphics[width=\columnwidth]{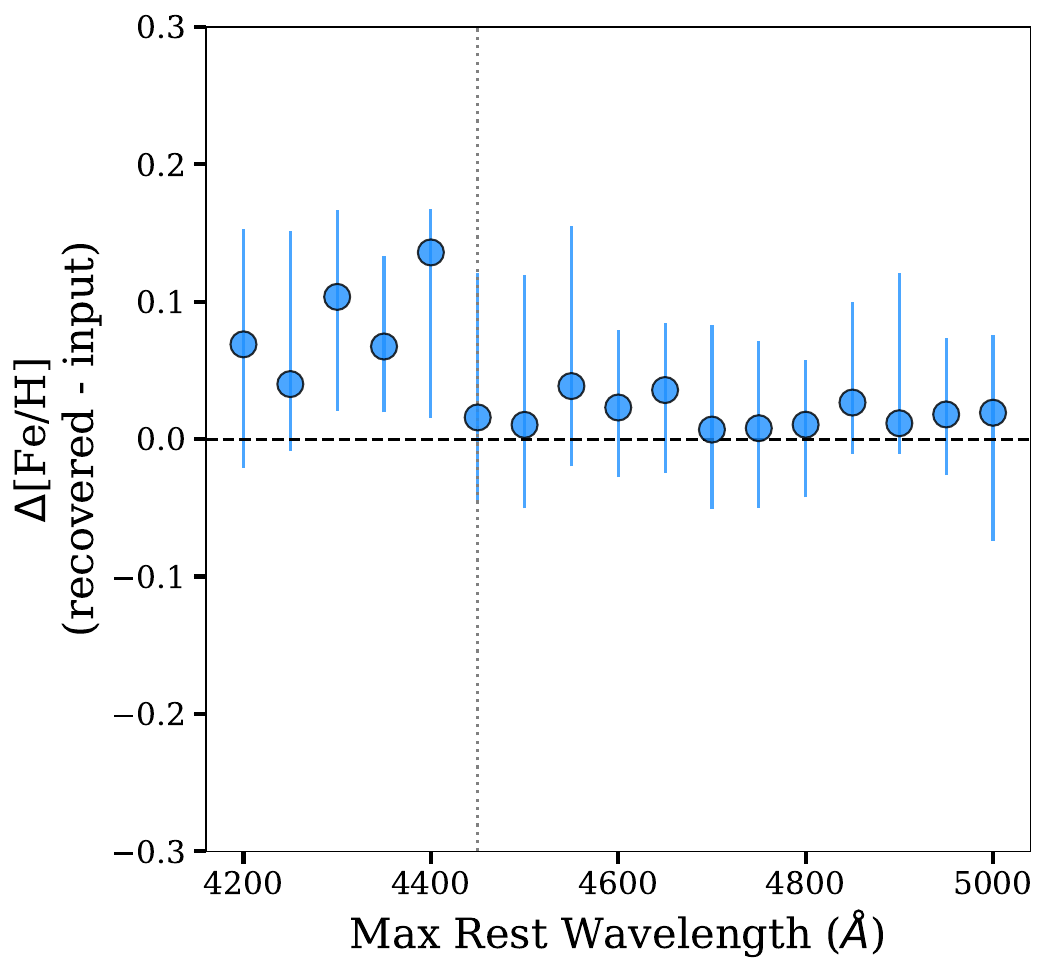}
    \caption{Mock spectra tests to determine the minimum maximum wavelength that can be used to reliably measure metallicity gradients.  This minimum maximum wavelength cut-off (4450 \AA) is indicated by the vertical dotted line (i.e. spectra need a maximum wavelength of at least 4450 \AA\ in order for us to recover metallicity gradients).  On the $x$-axis, we show the maximum simulated wavelength.  On the $y$-axis, we show the difference between the recovered and input [Fe/H].  Mock spectra are given a median S/N of 20 \AA$^{-1}$ (the S/N limit of our study).}
    \label{fig:paper_mockwave}
\end{figure}

We perform several tests with simulated mock LEGA-C spectra as discussed in Section~\ref{sec:sample_selection}.  In general, we make use of the \texttt{write\_a\_model} framework in \textsc{alf} to generate mock spectra.  We give the mock spectra specific values of $\sigma$, age, and elemental abundances, as well as an S/N of 1000 \AA$^{-1}$.  We then scale the mocks to different S/N.  To do this, we take the noise spectra from real LEGA-C galaxies and scale the noise to the desired S/N.  We also randomly sample the noise from a Gaussian distribution and add this random noise to our mock flux to achieve a more realistic simulated spectrum.

First, we test the S/N that we need to recover ages, [Fe/H], and Mg abundances in spatially resolved bins by generating several sets of mock spectra with typical properties (solar abundances, $\sigma = 150$ km s$^{-1}$, and an age of 3 Gyr) at varying levels of S/N.  For each S/N bin, we generate 40 mock spectra and fit each with \textsc{alf}, leaving age, [Fe/H], and [Mg/H] free and fixing all other abundances to their input values (similar to our core and outskirt fits for our real data).  We take the differences between the recovered and input parameters from each fit.  We compute the medians of all of these differences and plot the median and $1\sigma$ errors of the distribution of medians in Fig.~\ref{fig:SN_mocks}.  From this test, we determine that we need S/N $\gtrsim20$ \AA$^{-1}$ to reliably recover the parameters of interest.  \cite{Choi_2014} performed similar tests and came to a similar conclusion for quiescent galaxies at low-$z$.

We also use these mocks to test which elemental abundances we are able to recover by fixing different combinations of parameters, including different elements and velocity dispersion.  We find that our results are robust when we fix all elemental abundances except for Fe and sometimes Mg (depending on the wavelength range, see Section~\ref{sec:sample_selection}), and leave age and velocity dispersion free.

We note that there is a very small offset between the input and recovered ages in the leftmost panel of Fig.~\ref{fig:SN_mocks}.  We examine this by stacking all 40 of the MCMC age chains for the S/N $ = 20$ \AA$^{-1}$ test and plotting the distribution in Fig.~\ref{fig:mock_age_test}.  The distribution has a long tail towards old ages, which skews the median toward a slightly older age.  However, the peak of the distribution is precisely at the input age, which shows that \textsc{alf} is indeed recovering the input age.  We still use the median age in our analysis as the offset is very small.  Additionally, we note that because the ages are systematically offset to slightly older ages, this does not affect our conclusions as we examine the difference between the age in the core and the outskirt, and not the absolute value of the age.

Since the LEGA-C galaxies are at varying redshifts, they have variable rest-frame wavelength ranges, so we perform a second test to determine the shortest wavelength range for which we can still recover [Fe/H].  We similarly generate mock spectra as for the S/N tests, but with varying maximum wavelengths.  Specifically, we generate mock spectra with wavelengths ranging from $3700$ \AA $\leq \lambda \leq (4200-5000)$ \AA\ to supplement the real data.  We give the mock spectra $\sigma = 150$ km s$^{-1}$, an age of 3 Gyr, and a solar abundance pattern.  We generate 20 mock spectra for each wavelength range.  In this test, we treat our simulated spectra like our real integrated spectra and leave all parameters free in the \textsc{alf} fits.  We determine the differences between our recovered and input [Fe/H] and plot the medians in Fig.~\ref{fig:paper_mockwave} (similar to Fig.~\ref{fig:SN_mocks}).  We are able to reliably recover [Fe/H] down to a maximum wavelength of 4450 \AA.  We note that the accuracy of the age recovery is less dependent on the upper wavelength range, as there are many Balmer absorption lines redward of $3700\AA$. 

\begin{figure}
    \centering
    \includegraphics[width=\columnwidth]{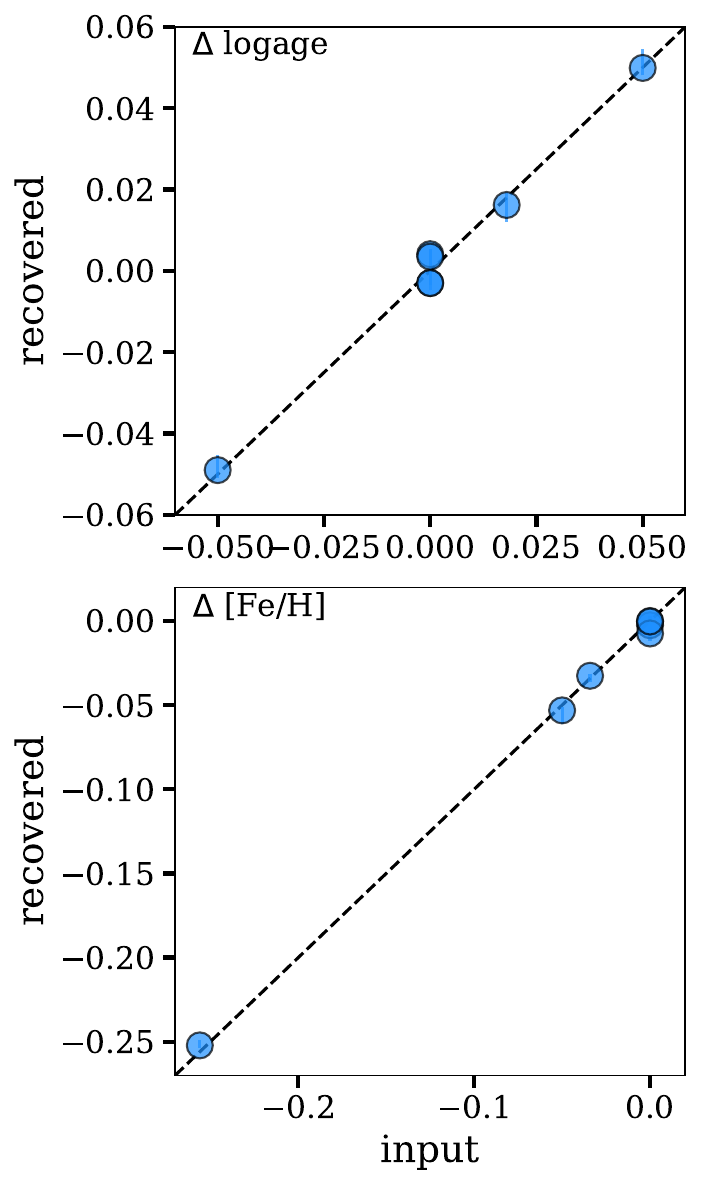}
    \caption{Mock spectra tests to demonstrate that we are able to recover age and metallicity gradients in our data.  The S/N of the simulated spectra is the same as that of the corresponding real LEGA-C spectra.  On the $x$-axes we show the input gradient and on the $y$-axes we show the recovered gradient.  The median gradients for $\log(\mathrm{age})$ are shown in the top panel and the median gradients for [Fe/H] are shown in the bottom panel.}
    \label{fig:paper_mockgrad}
\end{figure}

Finally, we note that there is an anticorrelation between our measured gradients in age and metallicity, so we assess whether this may be caused by the well-known age--metallicity degeneracy \citep{Worthey_1994, Bruzual_2003, Gallazzi_2005}. 
 We find that this is not the case.  In particular, even if the individual measured values of age and metallicity are affected by the degeneracy, the measured \textit{gradients} in these quantities should not change.  To test this, we generate several sets of 47 simulated LEGA-C galaxies with different combinations of age and metallicity gradients (i.e. no gradients, an age gradient but no metallicity gradient, a metallicity gradient but no age gradient, etc.).  We adopt the S/N of the corresponding real LEGA-C galaxies for each simulated set of spectra (ranging from a median value of between $\sim 20 - 113$ \AA$^{-1}$ in the simulated core spectra and $\sim 20 - 126$ \AA$^{-1}$ in the simulated outskirt spectra).  We find that we are always able to recover the input gradient within uncertainties.  We show this in Fig.~\ref{fig:paper_mockgrad}, where the comparison between our median recovered and input age gradients are shown in the top panel and the same for metallicity is shown in the bottom panel. 

\section{Duplicate objects}\label{sec:duplicates}
\begin{figure}
    \centering
    \includegraphics[width=\columnwidth]{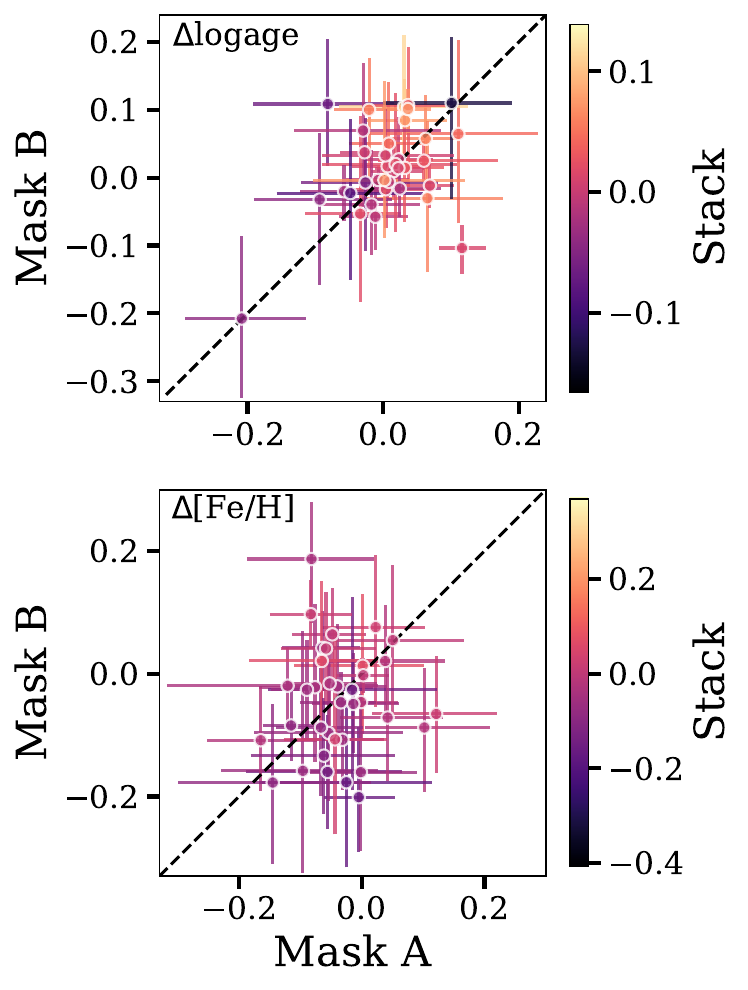}
    \caption{A comparison of the duplicate objects that are observed twice (in different masks) in LEGA-C. 
 The differences between the outskirt and core ages are shown in the top panel and the differences between the outskirt and core [Fe/H] values are shown in the bottom panel.  The results from one mask (Mask `A') are shown on the $x$-axis and the results from the other mask (Mask `B') are shown on the $y$-axis.  The symbols in each panel are colour coded by the stacked results.}
    \label{fig:paper_duplicates}
\end{figure}

There are several galaxies in LEGA-C that have been observed twice, in two different masks.  In the main body of the paper, we have completed our analysis treating each duplicate observation as an individual galaxy.  Here, we compare the fits to the galaxies in one mask to those in the other mask.  Here we call them Mask `A' and Mask `B', which should not be confused for LEGA-C's actual masks (numbered from 1 to 102).  We also compare our results to results from a fit to the stack of both duplicate observations.  We create these stacks for the integrated, core, and outskirt regions.  We do not smooth the spectra to the same resolution, as they are spectra of the same object.  We regrid the spectra to the same wavelength range (the different LEGA-C masks have slightly different wavelength coverage) using \textsc{spectres} \citep{SPECTRES} and mean stack the spectra.  In the stacked core and outskirts, we fix the abundances other than Fe and Mg to the values from the fits to the stacked integrated spectra.  In Fig.~\ref{fig:paper_duplicates}, we plot the difference in ages between the outskirt and core regions in the top panel and the difference in [Fe/H] between the outskirt and core regions in the bottom panel.  We compare the results from Mask A and Mask B and colour code by the stacked results.  In general, the results between all three fits are consistent within error bars.  This is an independent check of the robustness of the results. 

%%%%%%%%%%%%%%%%%%%%%%%%%%%%%%%%%%%%%%%%%%%%%%%%%%

% Don't change these lines
\bsp	% typesetting comment
\label{lastpage}
\end{document}